%% file: main.tex
\documentclass[onecolumn,notitlepage,superscriptaddress, nofootinbib,nobibnotes, aps,prd,10pt,preprintnumbers]{revtex4-2}

\usepackage{CJKutf8,enumitem}
\usepackage[T1]{fontenc}
\usepackage[utf8]{inputenc}
\usepackage{amssymb}
\usepackage{amsmath}
\usepackage{amsfonts}
\usepackage{mathrsfs}
\usepackage{amsthm}
\usepackage{cancel}
\usepackage[colorlinks=true,
  linkcolor=blue,
  citecolor=blue,
  urlcolor=blue]{hyperref}
\usepackage{xcolor}
\usepackage{tikz}
\usepackage{physics,xparse}

\usepackage[skip=2pt plus1pt, indent=10pt]{parskip}

\usetikzlibrary{arrows.meta,calc,positioning}
\definecolor{ink}{RGB}{38,58,80}
\definecolor{head}{RGB}{56,80,100}
\definecolor{fib}{RGB}{125,160,210}
\definecolor{fibd}{RGB}{105,140,195}
\definecolor{disc}{RGB}{210,216,232}
\definecolor{band}{RGB}{225,228,242}
\definecolor{plane}{RGB}{176,188,200}
\definecolor{planeb}{RGB}{160,190,190}
\definecolor{grid}{RGB}{205,214,226}
\definecolor{arr}{RGB}{80,105,125}
\definecolor{tl}{RGB}{0,150,150}
\definecolor{cy}{RGB}{140,245,245}
\definecolor{pk}{RGB}{217,70,176}
\definecolor{nn}{RGB}{50,90,120}
\definecolor{og}{RGB}{242,125,40}
\definecolor{pe}{RGB}{255,184,134}
\definecolor{pebg}{RGB}{255,240,232}

\let\tempone\itemize
\let\temptwo\enditemize
\renewenvironment{itemize}{\tempone\addtolength{\itemsep}{0.5\baselineskip}}{\temptwo}

\DeclareFontFamily{U}{mathx}{}
\DeclareFontShape{U}{mathx}{m}{n}{<-> mathx10}{}
\DeclareSymbolFont{mathx}{U}{mathx}{m}{n}
\DeclareMathAccent{\widehat}{0}{mathx}{"70}
\DeclareMathAccent{\widecheck}{0}{mathx}{"71}

\def\be#1\ee{\begin{align}#1\end{align}}
\def\bg#1\eg{\begin{gather}#1\end{gather}}
\def\bsub#1\esub{\begin{subequations}#1\end{subequations}}

\def\cE{\mathcal{E}}

\def\cH{\mathcal{H}}
\def\cL{\mathcal{L}}
\def\cM{\mathcal{M}}

\def\cO{\mathcal{O}}

\def\cR{\mathcal{R}}
\def\cS{\mathcal{S}}

\def\dr{{\rm dr}}

\newcommand{\f}{\frac}
\def\d{\mathrm{d}}
\def\p{\partial}

\definecolor{darkpastelgreen}{rgb}{0.01,0.75,0.24}
\definecolor{revisionblue}{rgb}{0.05,0.20,0.75}

\begin{document}

\title{
{Relational Cosmology:
Covariant and Canonical Dressings,\\
and the Hamiltonian of Internal Time }}

\author{Francesco Sartini} 
\email{francesco.sartini@oist.jp}
\affiliation{Qubits and Spacetime Unit, Okinawa Institute of Science and Technology Graduate University (\begin{CJK}{UTF8}{min}沖縄科学技術大学院大学\end{CJK}), Onna, Okinawa 904 0495, Japan}

\date{\today}

\begin{abstract}
We compare covariant and canonical relational dressings with gauge-fixing reductions for the flat FLRW minisuperspace with a homogeneous massless scalar. We distinguish covariance of the dressing map from symmetries of the relational theory: fixing a clock makes its frame fields background structures, so its reorientations act as vector fields on relational field space rather than as spacetime diffeomorphisms. This loss of relational-time reparametrization covariance permits a nontrivial relational Hamiltonian, distinct from the undressed and dressed Hamiltonian constraints, which vanish on the constraint surface. We identify the conditions under which the dressed equations reproduce the original constrained dynamics. The relational generator defines a Hamiltonian flow even when relational-time translations are not symmetries of the dressed action. In each clock frame, it corresponds to the canonical complete observable and reduced Hamiltonian on the appropriate solution space, through the Legendre map and relational evaluation. The classical descriptions are compatible but retain distinct variational principles and intermediate phase spaces; their quantum equivalence remains an open question.
\end{abstract}

\maketitle

\tableofcontents

\section{Introduction}

General relativity is a background-independent theory whose dynamical content is constrained by spacetime diffeomorphism invariance. In the canonical formulation, this invariance is encoded in the Hamiltonian and momentum constraints. This gives rise to several intertwined aspects of the problem of time: the construction of physical observables, the recovery of a notion of physical dynamics, and the possible dependence of a reduced description on the choice of clock \cite{Kuchar:1991qf,Isham:1992ms,Anderson:2010xm,Anderson:2017jij,Rovelli:2004tv}.

The construction of observables requires quantities invariant under the transformations generated by the constraints. But, unlike in ordinary gauge theories, where the Hamiltonian generally generates evolution independently of the internal gauge transformations, in general relativity the Hamiltonian is itself a constraint, and its flow is part of the gauge redundancy. 
Consequently, the identification of physical evolution requires specifying it relationally, rather than referring to an external time parameter.

Relational approaches address these problems by describing physical quantities through correlations between dynamical variables. The foundations of the observable problem in general relativity go back to early analyses of constrained Hamiltonian systems \cite{dirac_lectures_2001,Bergmann:1961wa}, while the construction of partial and complete observables was subsequently developed into a systematic framework \cite{Rovelli:1990ph,Rovelli:1990pi,Rovelli:2001bz,Dittrich:2004cb,Dittrich:2005kc,Dittrich:2006ee,Dittrich:2007jx,Tambornino:2011vg,Thiemann:2007zz}. In cosmology, matter fields and geometrical variables can serve as internal clocks on suitable regions of the solution space. Dust reference systems provide a prominent example \cite{Brown:1994py}, while scalar-field clocks have been extensively studied in minisuperspace \cite{Blyth:1975is,Hawking:1983hn,Ashtekar:2006uz,Ashtekar:2007em,Kamenshchik:2018pwf,Gielen:2020abd,Kiefer:2021zdq}.

However, relational evolution is not uniquely determined by the underlying constrained dynamics: it depends on the choice of clock, and such a choice is not innocuous. Different choices can lead to distinct semiclassical descriptions \cite{Bojowald:2010xp,Bojowald:2010qw,Hohn:2011us} and, at the quantum level, to potentially inequivalent theories, differing in their physical inner product, self-adjointness, unitarity, and interpretation of singularity resolution \cite{Kuchar:1991qf,Isham:1992ms,Hartle:1995vj,Malkiewicz:2014fja,Malkiewicz:2015fqa,Malkiewicz:2016hjr,Malkiewicz:2017cuw,Gielen:2020abd,Fragolino:2026qij}. This is tied to the fact that quantizing before or after reducing the constraints are, in general, inequivalent procedures \cite{Ashtekar:1982wv,Ashtekar:1991hf,Schleich:1990gd,Kunstatter:1991ds,Hajicek:1990eu,Romano:1989zb,Loll:1990rx,Kaminski:2009qb,Domagala:2010bm,Giesel:2016gxq,Thiemann:2004wk,Dittrich:2015vfa}. The comparison between Dirac, reduced, and Page--Wootters quantizations \cite{Page:1983uc,Wootters:1984wfv,Giovannetti:2015qha,Gambini:2008ke} has been investigated using the language of quantum reference frames (QRFs) \cite{Hoehn:2019fsy,Hohn:2018iwn,Hoehn:2020epv}. More generally, the QRF perspective treats clocks and reference systems as dynamical degrees of freedom and relates descriptions associated with different internal perspectives \cite{Giacomini:2017zju,Vanrietvelde:2018pgb,Vanrietvelde:2018dit,delaHamette:2020dyi}.

These questions are not restricted to the problem of time. In field theory and gravity, the choice of relational frame is also tied to the construction of observables and, in particular, to the problem of defining localized quantities in a diffeomorphism-invariant manner. Gravitational dressings have a substantial history in the construction of diffeomorphism-invariant observables \cite{Donnelly:2015hta,Donnelly:2016rvo,Goeller:2022rsx}. Closely related ideas appear in the literature on edge modes, where boundary variables can likewise be interpreted in terms of dynamical reference frames \cite{Carrozza:2021gju,Carrozza:2022xut,Donnelly:2016auv,Freidel:2020xyx}.

A complementary covariant approach constructs relational dressed observables by using dynamical reference fields at the Lagrangian level. The covariant dressing considered here follows \cite{Goeller:2022rsx,Carrozza:2022xut,Aguilar-Gutierrez:2026svf,AFHMtime}. When the frame is a set of scalars, the map from spacetime to relational spacetime is at once a set of (functionals of the) fields and a field-dependent diffeomorphism, and the dressed theory is obtained by pulling back the fields, the action, and the presymplectic structure along it.
The dressing aligns the frame, viewed as a coordinate on field space, with the gauge fibers. Each gauge-fixed leaf is then isomorphic to the space of gauge orbits, and the dressing can be regarded as equivalent to a complete gauge-fixing.
For this to be meaningful, the frame must be compatible with the solution spaces in the sense that they provide a non-degenerate set of coordinates on the on-shell (relational) spacetime.

The dressing, however, should not be conflated with a general gauge-fixing imposing a spacetime condition on the fields to render the equations of motion hyperbolic. The two coincide only if the gauge-fixing is complete, including in time. In general, gauge-fixing variables at the level of the action, before variation, can remove independent equations of the original theory \cite{Motohashi:2016prk}. 

The role of time makes the comparison between covariant dressing and the canonical description particularly delicate. The transformations generated by spacetime reparametrizations and the transformations generated by the Hamiltonian constraint are related, but their algebras do not coincide. This distinction is already present in the minisuperspace model considered here: time reparametrizations form a non-abelian Lie algebra, whereas the Poisson algebra of the single Hamiltonian constraint is abelian. The relation between the two becomes clear in the extended canonical formulation, where the lapse is included as a canonical variable together with its conjugate momentum. The generator of time reparametrizations then involves both the primary and secondary constraints, with parameters related by a field-dependent redefinition involving the lapse. Consequently, the algebra of the Hamiltonian constraint alone does not capture the full action of time reparametrizations, in particular its action on the lapse \cite{Pons:1996av,Pons:2003uc}. This distinction is important when comparing covariant, canonical and gauge-fixed constructions.

Here we reverse the order of the construction: we start from the covariant description, where spacetime reparametrizations act on all the fields, and compare it with the canonical picture only afterwards. For a non-degenerate dressing, the slicing is specified relationally by the frame coordinate $u$, which becomes a background coordinate for the remaining fields in the dressed description. This removes the residual time-reparametrization gauge freedom and allows for a canonical Hamiltonian description with respect to $u$. It does not, however, eliminate the constraint inherited from the original theory: the dressed lapse retains its role in the variational principle, and its equation of motion imposes the covariantly dressed Hamiltonian constraint. In the extended canonical description, this constraint is accompanied by the primary constraint associated with the lapse momentum. Unlike the original Hamiltonian constraint, however, the dressed constraint does not generate gauge transformations in the relational theory; it fixes the time evolution of the lapse itself. The resulting relational Hamiltonian must therefore be distinguished both from the dressed constraint and from the canonical generator of time reparametrizations in the original theory. Establishing the precise correspondence between these structures is an important part of the analysis below.

This paper aims to examine these relations in the simplest gravitational setting in which they can be compared explicitly: spatially flat homogeneous cosmological minisuperspace with a minimally coupled massless scalar field \cite{Blyth:1975is,Hawking:1983hn,Kamenshchik:2018pwf,Ashtekar:2006uz,Ashtekar:2007em,Ashtekar:2011ni,Banerjee:2011qu,Oriti:2016qtz,Gielen:2016uft,Gielen:2018fqv,Hohn:2018iwn}. The model retains the essential ingredients of a constrained gravitational system, including a lapse enforcing the Hamiltonian constraint and dynamical variables that can serve as internal clocks, while avoiding non-local the need for non-local dressing fields. It also provides a simple classical example for the more general construction of relational Hamiltonians described in \cite{AFHMtime}.

We consider several candidate frames and compare the resulting relational descriptions. In the canonical construction, a suitable clock variable must vary along the gauge orbits generated by the Hamiltonian constraint; locally, this requires a non-vanishing Poisson bracket with the constraint on the region of interest \cite{Dittrich:2004cb,Dittrich:2005kc}. At the covariant level, the corresponding requirement is not simply that the frame be a spacetime scalar. Rather, the frame map must be compatible with the solution space and define a non-degenerate coordinate along the relevant gauge orbits. Equivalently, the dressed variational principle must retain enough independent equations to recover the dressing of the solution space of the original constrained dynamics.

Our contribution is an explicit comparison of covariant dressing and its associated variational principle, with both canonical dressing and gauge-fixed reduction. We show how fixing the lapse at the level of the action generally differs from introducing a dynamical reference frame, and compare the resulting equations of motion and presymplectic structures. We identify the generator of relational evolution and distinguish it from a conserved Noether charge, which arises only when the relational theory possesses an additional symmetry, as in the “shift symmetry'' associated with a scalar-field clock. For both the scalar-field and scale-factor frames, we establish a classical correspondence between the covariant and canonical descriptions, at the level of the equations of motion and the physical symplectic structure on the relevant solution spaces, despite the inequivalence of their variational principles off-shell. In the massless-scalar model, the two choices of clock describe the same relevant classical solutions, although they lead to different relational descriptions. We do not establish an equivalence between the corresponding quantum theories, which would require a separate analysis of their quantization.

The model thus provides a controlled setting in which to disentangle covariant dressing, canonical constraint reduction, and relational Hamiltonian evolution. We begin by introducing the FLRW model and the dressing prescription in Sec.~\ref{sec:flrw_dressing}. We then investigate the covariance of the dressed theory and construct its relational Hamiltonian for a specific choice of frame in Sec.~\ref{sec:cov_rel_Ham}. In Sec.~\ref{sec:canonical}, we develop the corresponding Dirac constraint analysis and canonical reduction, establishing the relation between the covariant and canonical descriptions. Throughout, we examine the generator of relational time evolution and its connection with the Page--Wootters perspective.

\section{FLRW model and covariant dressing}
\label{sec:flrw_dressing}

We consider a spatially flat, homogeneous minisuperspace coupled to a homogeneous, massless scalar field to illustrate the construction of a covariant dressing and the corresponding relational dynamics. Minisuperspace models truncate the infinitely many degrees of freedom of general relativity to a finite set of homogeneous variables. Such truncations can arise as controlled approximations in particular regimes, for example in long-wavelength or gradient expansions in cosmology, or in velocity-dominated (BKL-type) regimes near spacelike singularities.
Here, our interest is instead primarily conceptual: minisuperspace provides a simple toy model of a one-dimensional diffeomorphism-invariant theory, retaining the Hamiltonian constraint and time-reparametrization invariance while allowing their interplay with relational clocks and variational dynamics to be studied explicitly.

We restrict to the Friedmann--Lemaître--Robertson--Walker (FLRW) class of metrics
\begin{equation}
\d s^2=-N(t)^2\,\d t^2+a(t)^2\,\d\mathbf{x}^2,
\end{equation}
where $N$ is the lapse and $\d\mathbf{x}^2$ is the flat spatial metric. For the noncompact spatial slices, we introduce a fiducial cell and absorb its coordinate volume into the normalization of the reduced action; this serves only to regularize the spatial integral and does not introduce a physical boundary. After integrating over the fiducial cell, the resulting theory is a mechanical system (a $1+0$-dimensional field theory), with all spatial dependence removed.

The Einstein--Hilbert action, minimally coupled to a massless scalar $\Phi$, then reduces to
\begin{equation}
\label{eq:FLRW_action}
\cS=\int_{\mathbb R}\d t\left[
\frac{a^3\dot\Phi^2}{2N}
-\frac{3}{8\pi G}\frac{a\,\dot a^2}{N}
\right].
\end{equation}
The corresponding Euler--Lagrange equations are
\begin{align}
\cE_a
&=\frac{3}{8\pi G}\left(
\frac{\dot a^2}{N}
+\frac{4\pi G}{N}a^2\dot\Phi^2
+2a\,\partial_t\!\left(\frac{\dot a}{N}\right)
\right), \notag\\
\cE_\Phi
&=-\frac{\d}{\d t}\left(\frac{a^3\dot\Phi}{N}\right),
\label{eq:FLRW_eoms}\\
\cE_N
&=\frac{3}{8\pi G}\frac{a\,\dot a^2}{N^2}
-\frac{a^3\dot\Phi^2}{2N^2}. \notag
\end{align}

These equations encode the homogeneous and isotropic Einstein--scalar dynamics. Variation with respect to the lapse, $\cE_N \approx 0$, gives the first Friedmann equation\footnote{We use the notation $\approx$ for on-shell equality}, which is the Hamiltonian constraint at a canonical level. Variation with respect to the scale factor, $\cE_a\approx0$, gives the acceleration equation (the second Friedmann equation), while
$\cE_\Phi\approx0$ is the homogeneous Klein--Gordon equation on FLRW background. In particular, the latter is the continuity equation for the homogeneous scalar-field energy density.

It is important to distinguish the symmetry reduction leading to \eqref{eq:FLRW_action} from the gauge reduction considered below. The homogeneous and isotropic ansatz restricts the space of field configurations of general relativity to a symmetry-invariant sector. In the present model, this restriction can consistently be implemented at the level of the action, yielding the corresponding symmetry-reduced Einstein equations. This truncation removes the inhomogeneous degrees of freedom but does not amount to choosing a representative along a gauge fiber\footnote{Spatial diffeomorphisms act on the non-dynamical fields within the homogeneous sector and do not give rise to additional dynamical gauge degrees of freedom in the present model.}. Within the resulting minisuperspace, only time reparametrizations remain as a gauge redundancy. Our analysis therefore proceeds by truncating general relativity to its homogeneous and isotropic sector before constructing relational observables. The two operations need not commute: constructing observables in the full theory and subsequently imposing homogeneity may retain information about inhomogeneous degrees of freedom that is absent when truncation precedes dressing. We do not investigate this comparison here, and all results refer to the minisuperspace theory obtained by truncating first.

In this work, ``covariant'' refers specifically to this residual one-dimensional diffeomorphism invariance, rather than to the full spacetime diffeomorphism group. Accordingly, we use the language of covariant phase space to describe the space of histories and their variations. The variation of the Lagrangian one-form can be written as
\begin{equation}
\delta L
=
\left(\cE_a\delta a+\cE_\Phi\delta\Phi+\cE_N\delta N\right)\d t 
+\d\theta : = {\rm Eom} +\d \theta\,,
\end{equation}
where the pre-symplectic potential is the spacetime zero-form
\begin{equation}
\label{eq:FLRW_theta}
\theta
=\frac{a^3\dot\Phi}{N}\,\delta\Phi
-\frac{3a\,\dot a}{4\pi GN}\,\delta a
=:\pi_\Phi\,\delta\Phi+\pi_a\,\delta a.
\end{equation}
Thus
\begin{equation}
\pi_\Phi:=\frac{a^3\dot\Phi}{N},
\qquad
\pi_a:=-\frac{3a\,\dot a}{4\pi GN}.
\end{equation}
Here $\theta$ is a spacetime zero-form and a one-form on field space. The “covariant momenta'' $\pi_a$ and $\pi_\Phi$ are introduced as the coefficients of the field-space variations in $\theta$; at this covariant stage, they are functions of the configuration fields and their velocities, not additional independent variables. The time diffeomorphism symmetry acts as
\begin{equation}
\label{eq:FLRW_gauge_transf}
\delta_\xi a=\xi\dot a,\qquad
\delta_\xi\Phi=\xi\dot\Phi,\qquad
\delta_\xi N=\partial_t(\xi N).
\end{equation}
The associated Noether identity is
\begin{equation}
\label{eq:FLRW_relationship}
N\dot{\cE}_N=\dot\Phi\,\cE_\Phi+\dot a\,\cE_a.
\end{equation}
This is the minisuperspace counterpart of the contracted Bianchi identity: it expresses the dependence among the Einstein equations that follows from diffeomorphism invariance. In particular, it relates the evolution of the constraint to the scale-factor and scalar equations. This is an off-shell identity among the Euler--Lagrange equations.

The general solutions make the role of time-reparametrization invariance explicit. They consist of two branches, distinguished by whether $\dot a$ and $\dot\Phi$ have the same or opposite signs. Both branches can be parametrized in terms of an arbitrary monotonic function $f(t)$ as
\begin{equation}
\label{eq:FLRW_general_solutions}
N(t)\approx e^{f(t)} \dot f(t),\qquad
a(t)\approx a_0 e^{f(t)/3},\qquad
\Phi(t)\approx \Phi_0\pm\frac{f(t)}{\sqrt{12\pi G}}.
\end{equation}
Here $a_0$ and $\Phi_0$ are integration constants, while the arbitrary function $f$ encodes the freedom to choose the time parametrization. Choosing $f(t)=t$ selects a convenient representative of each branch. The two signs distinguish solutions in which the scalar field increases or decreases with proper time. On either branch, both $\dot a$ and $\dot\Phi$ are non-vanishing.
Consequently, both $a$ and $\Phi$ provide admissible relational clocks on these intervals.\footnote{We set aside the degenerate configurations with $f=\mathrm{const.}$, for which $N=0$ and the metric is degenerate. Although they arise as a formal constant-$f$ limit of the parametrization above, they lie outside the regular configuration space on which the original action and its Euler--Lagrange equations, containing inverse powers of $N$, are defined. They are therefore excluded from the present classical solution space. Their exclusion may nevertheless matter in a quantum treatment: it removes a degenerate sector that lies between, or at the common boundary of, the two non-degenerate branches, depending on the topology assigned to the configuration space. In particular, it may affect the integration domain and admissible contour deformations in the lapse path integral, and hence the resulting propagator \cite{Mottola:1995sj,Halliwell:1988wc}. We do not discuss this quantum effect here.}

It is worth emphasizing that the gauge parameter $\xi(t)$ in \eqref{eq:FLRW_gauge_transf} is an arbitrary function, and that the corresponding transformations form the Lie algebra of vector fields on $\mathbb R$:
\begin{equation}
[\xi_1\partial_t,\xi_2\partial_t]
=(\xi_1\dot\xi_2-\xi_2\dot\xi_1)\partial_t.
\end{equation}
This is the algebra of time reparametrizations in the reduced mechanical model. It is important, however, not to identify this algebra directly with the Poisson algebra of the Hamiltonian constraint.

The distinction is already visible in the canonical formulation before any relational gauge is chosen. The minisuperspace contains a single Hamiltonian constraint $\cH\approx0$, whose Poisson bracket with itself vanishes identically, $\{\cH,\cH\}=0$. Its constraint algebra is therefore abelian. The algebra of time reparametrizations, on the other hand, is the non-abelian Lie algebra displayed above. This is not simply the familiar distinction between spacetime diffeomorphisms and hypersurface deformations that arises when passing from a covariant description to ADM variables. Even after restricting to homogeneous transformations that preserve the spacelike character of the slices, the two algebras differ.

The relation between them involves the lapse, and it is better described in the context of the extended phase space, where one includes the canonical pair $(N,\pi_N)$ and the primary constraint $\pi_N\approx0$, in addition to the secondary constraint $\cH\approx0$. The generator of gauge transformations is then a combination of both constraints, with a time-dependent parameter and its derivative. In the notation used below,
\begin{equation}
\label{eq:FLRW_extended_gauge_generator_intro}
G_\epsilon=\dot\epsilon\,\pi_N+\epsilon\,\cH.
\end{equation}
generates the transformations \eqref{eq:FLRW_gauge_transf} when $\epsilon=N\xi$. Thus, the non-abelian reparametrization algebra is represented on the extended phase space through a field-dependent relation between the vector field $\xi$ and the parameter $\epsilon$ multiplying the canonical constraints. The distinction is essential: the algebra of the secondary constraint alone does not encode the transformation of the lapse. We return to this point in the canonical analysis, where the extended phase space and the full generator can be treated explicitly.

Because $\epsilon=N\xi$ is field-dependent, the correspondence between the two descriptions also involves the adjusted composition law for gauge transformations with field-dependent parameters. Thus, the abelian Poisson algebra of the constraints
with field-independent $\epsilon$ is not in conflict with the non-abelian Lie algebra of time reparametrizations. The latter is recovered
only after accounting for the field dependence of the map between them.

\subsection{Dressing and the variational principle}

A covariant dressing constructs relational fields by specifying their values with respect to a chosen frame variable $R$, itself a functional of the dynamical fields \cite{Carrozza:2022xut,Aguilar-Gutierrez:2026svf,AFHMtime,Goeller:2022rsx}. We use DeWitt's condensed notation, in which $I$ collectively labels the field species and their components; in the present model, $\psi_I\in\{a,\Phi,N\}$. The frame defines a map from spacetime to a relational spacetime,
\begin{equation}
R:\cM\longrightarrow\mathfrak m,
\end{equation}
which we assume to be invertible on the region under consideration. The dressed fields are then obtained by transporting the original fields to relational spacetime. For covariant tensors and differential forms, this is achieved by pulling back along $R^{-1}$:
\begin{equation}
\psi_I^\dr={R^{-1}}^*[\psi_I]
:=\cR[\psi_I].
\end{equation}
Thus, $\cR$ expresses the fields in coordinates adapted to the chosen frame. Contravariant tensors and vector fields are instead transported by the pushforward along $R$. For notational convenience, we denote both operations by $\cR$
\begin{equation}
\xi^\dr=R_*[\xi]:=\cR[\xi],
\end{equation}
with the appropriate pullback or pushforward understood from the tensorial character of the field. In particular, the lapse transforms
as the component of the one-form $N\,\mathrm dt$, rather than as a spacetime scalar.

In the present minisuperspace model, the spacetime manifold is initially $\cM=\mathbb R$, with coordinate $t$, and the gauge
redundancy consists of time reparametrizations. A single frame map therefore suffices to define relational time. We denote its coordinate by $u$ and write the dressed fields explicitly as
\begin{equation}
\label{eq:FLRW_dressed_observable}
\psi_I^\dr(u)
=
\psi_I(t)\big|_{R(t)=u}.
\end{equation}
That is, the dressed field at $u$ is the value of the original field at the spacetime point where the frame takes the value $u$. Throughout, a dot $\dot\,$ denotes differentiation with respect to the time coordinate appropriate to the field: $t$ for undressed fields and $u$ for dressed fields.

The construction and its general properties are summarized in App.~\ref{app:covariant phase space} and in Refs.~\cite{Carrozza:2022xut,Goeller:2022rsx,Aguilar-Gutierrez:2026svf,AFHMtime}. A dressing should be distinguished from gauge-fixing. The former constructs relational fields by expressing the original fields with respect to a field-dependent frame; the latter imposes conditions that select representatives along gauge fibers. A sufficiently regular and complete dressing locally provides a parametrization equivalent to choosing a gauge-fixing section.

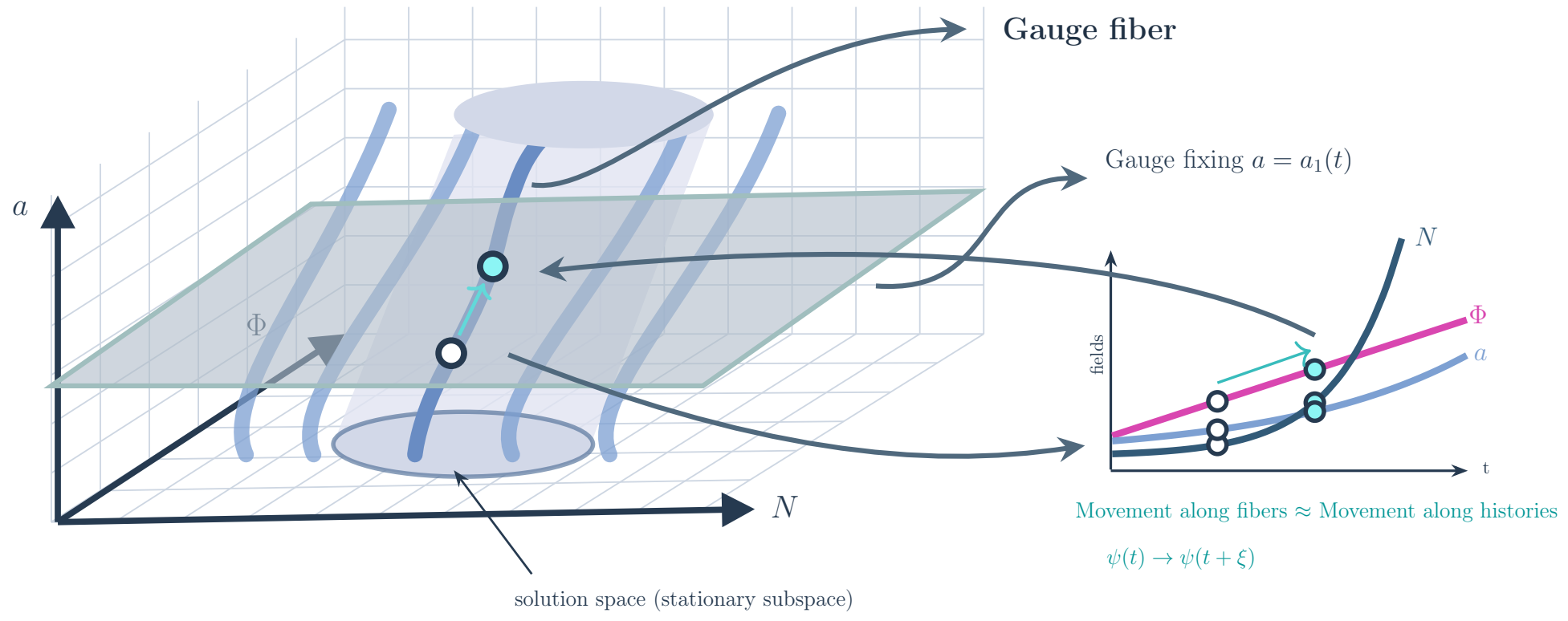
\begin{figure}[t]
\centering
 \resizebox{0.8\textwidth}{!}{%
\input{fieldspace_original}
}
\caption{
Schematic representation of field space. A gauge-fixing section
selects representatives along gauge fibers. The solution space is the subspace of stationary histories (the ones that locally extremise the action functional $\delta S \approx 0 \Leftrightarrow \cE_I \approx 0$). For illustration purposes, it is represented as a “tube'' in the figure above, even though it is a hypersurface (in the infinite-dimensional history space). The time evolution within a history is equivalent to a “vertical'' translation along a direction within the fibers
}
\label{fig:gauge_section}
\end{figure}

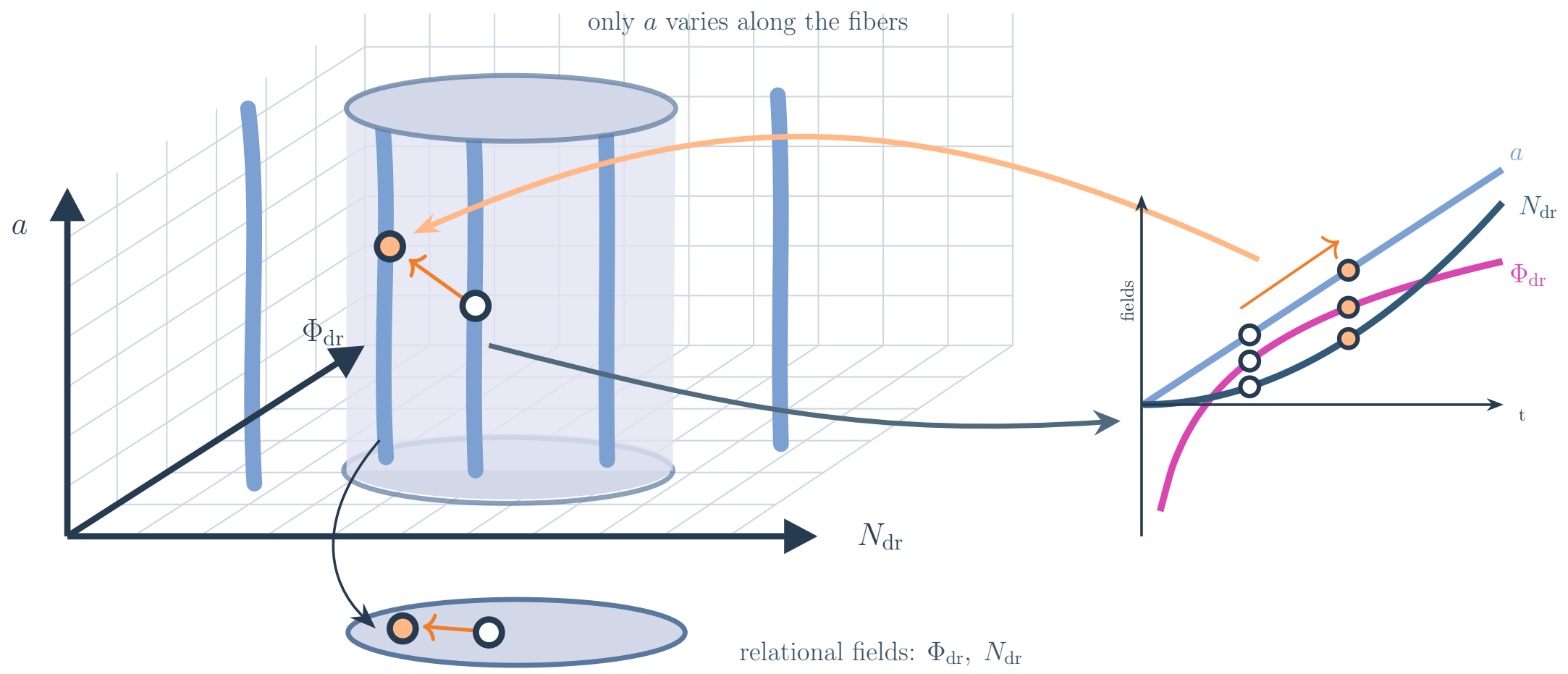
\begin{figure}[t]
\centering
 \resizebox{0.8\textwidth}{!}{%
\input{fieldspace_dressed}
}
\caption{
A complete dressing
provides relational coordinates that locally parametrize the same
gauge-invariant information. The correspondence with a
gauge-fixing section requires the frame to be non-degenerate along
the fibers and the section to intersect them transversally.
Schematic representation of the relationally dressed field space. The frame defines relational coordinates along the gauge fibers, while the remaining dressed fields describe the corresponding relational data. The construction is locally equivalent to a gauge section only where the frame map is non-degenerate and intersects each relevant orbit uniquely. The relational time evolution flow follows a history in the dressed solution space, changing the dressed observables. Mapped back to
field space, it combines horizontal motion between
fibers with vertical gauge motion along them.}
\label{fig:dressed_field}
\end{figure}

To make the distinction between dressing and gauge-fixing precise, consider the space of field configurations and its quotient by spacetime diffeomorphisms. The diffeomorphism orbits are the fibers of the projection from field space to the space of gauge-equivalence classes. A gauge-fixing section selects one representative on each fibers, at least locally, whereas a dressing uses a field-dependent coordinate map to construct relational fields. Figures~\ref{fig:gauge_section} and~\ref{fig:dressed_field} illustrate these two descriptions.

However, this is a kinematical correspondence: it does not, by itself, establish equivalence between the gauge-fixings of the solution space and the solution space of the dressed theory.
The relevant question is therefore not merely whether a candidate frame transforms covariantly, but whether it defines a non-degenerate
coordinate along the gauge fibers on the solution space of interest, and whether the resulting dressed variational principle retains the
independent equations of the original theory. In particular, a frame that is constant on shell cannot serve as a relational time on that
branch. Conversely, imposing an incomplete gauge-fixing condition directly in the Lagrangian before variation can remove the Euler--Lagrange equation associated with the variable being fixed and thus should not be interpreted as equivalent to a dressing \cite{Motohashi:2016prk}.\\

\paragraph{The scalar field as a frame.}

We first consider the scalar field as the frame, $R=\Phi$. Setting $\Phi_\dr=u$ gives
\begin{equation}
\label{eq:FLRW_phi_dr_action}
\mathcal R_\Phi^*[L]
=\left(
\frac{a_\dr^3}{2N_\dr}
-\frac{3}{8\pi G}\frac{a_\dr\dot a_\dr^2}{N_\dr}
\right)\d u.
\end{equation}

The independent equations obtained by varying $a_\dr$ and $N_\dr$
are
\begin{align}
\cE_a^\dr
&=\frac{3}{8\pi G}\left(
\frac{\dot a_\dr^2}{N_\dr}
+\frac{4\pi G}{N_\dr}a_\dr^2
+2a_\dr\,\partial_u\!\left(\frac{\dot a_\dr}{N_\dr}\right)
\right), \notag\\
\cE^N_\dr
&=\frac{3}{8\pi G}\frac{a_\dr\dot a_\dr^2}{N_\dr^2}
-\frac{a_\dr^3}{2N_\dr^2}.
\label{eq:FLRW_phi_dr_eoms}
\end{align}

These are the dressed counterparts of the independent bare equations. The scalar equation is not varied independently because the frame condition has already fixed $\Phi_\dr=u$. Nevertheless, when $\cE_a^\dr$ and $\cE_N^\dr$ both vanish, the Noether identity associated with time-reparametrization invariance implies the remaining scalar equation, provided that $\dot\Phi\neq0$ on the solution branch under consideration, which is indeed the case on all the solution space considered here. If $\Phi$ were constant on shell, this inference would fail, since the Noether identity would no longer allow the scalar equation to be recovered from the other two.

The scalar clock therefore provides an example in which the frame is non-degenerate on the classically relevant solutions, and the dressed variational principle retains the independent dynamics. It also illustrates that covariance of the frame transformation, although necessary, is not sufficient to guarantee a valid relational description. The “momentum-frame'' examples below make this distinction particularly clear.\\

\paragraph{The lapse as a frame.}

The lapse provides a useful contrast, but it should not be regarded as a frame variable in the same sense as the scalar field. Since $N\,\mathrm dt$ is a one-form, the lapse itself is not a scalar whose value can be used to define a coordinate map $R:\cM\to\mathfrak m$. The condition $N=1$ instead fixes the time parametrization to proper time. It is a gauge condition on the original description, not a relational dressing by the lapse.
In particular, it fixes the local time-reparametrization freedom but leaves a residual freedom to translate the origin of proper time. The distinction is important because imposing this condition in the action before variation has important consequences.

To see this, consider the action obtained by expressing the fields
in proper time and imposing $N=1$ before varying:
\begin{equation}
\label{eq:FLRW_lapse_dr_action}
L|_{N=1}
=\left(
\frac{a^3\dot\Phi^2}{2}
-\frac{3}{8\pi G}a\,\dot a^2
\right)\d\tau,
\end{equation}
where $\tau$ is the proper-time coordinate. 

This gives a simple mechanical action, but the lapse is no longer varied, so its equation $\cE_N=0$ is absent. On solutions of the remaining equations, the Noether identity gives only $\dot{\cE}_N=0$, not $\cE_N=0$. Consequently, the lapse-fixed variational problem admits solutions with a nonzero constant value of $\cE_N$, unless the constraint is imposed separately on the initial data. 

There is an important qualification. Fixing the local lapse does not necessarily remove the total proper-time as a field from a path integral \cite{Halliwell:1988wc,Mottola:1995sj} (sometimes referred to as ''modulus''). If that modulus remains an integration variable and is integrated over, it can impose the constraint, as in proper-time formulations of constrained systems. The classical statement above concerns the variational problem in which the lapse is fixed and proper time is treated as a parameter; it should not be read as a claim that every proper-time path integral loses the constraint.

Accordingly, the proper-time condition is useful as an example of a partial gauge-fixing, but not of a complete relational dressing.
The residual translation and the loss of the lapse equation are separate issues: the former concerns the choice of time origin,
whereas the latter concerns which equations are retained by the variational principle.
\footnote{In the presence of boundaries, we could still leave some residual ambiguities in the gauge-fixing, if they can be \textit{cured} by boundary conditions (and not by restricting the initial condition), a point we leave to discussions in upcoming works \cite{JTproj}}.\\

\paragraph{Momentum as a frame.}
One can also consider a momentum as a frame variable. Although this is less intuitive from the configuration-space perspective, both momenta in \eqref{eq:FLRW_theta} transform as scalars under time reparametrizations:
\begin{equation}
\delta_\xi\pi_a= \f{\delta\,\pi_a}{\delta a } \delta_\xi a + \f{\delta\,\pi_a}{\delta \dot a} \p_t \delta_\xi a + \f{\delta \pi_N}{\delta N} \delta_\xi a  = \xi\dot\pi_a,\qquad
\delta_\xi\pi_\Phi=\xi\dot\pi_\Phi.
\end{equation}
The transformation law alone therefore does not distinguish configuration variables from momenta as candidate frames. Their admissibility instead depends on their behaviour along solutions.

It is important, however, to distinguish this Lagrangian use of the momentum functions from the canonical treatment developed
below. At this stage, $\pi_a$ and $\pi_\Phi$ are the functions of the configuration fields and their velocities defined in \eqref{eq:FLRW_theta}; they are not independent phase-space coordinates. Choosing $\pi_a$ as a frame therefore amounts to
imposing a condition on the corresponding combination of $a$, $\dot a$, and $N$. We solve this condition for the lapse and
substitute the result into the action, which is then varied with respect to the remaining dressed configuration fields. This
procedure is distinct from the canonical one, in which momenta are independent coordinates on the extended phase space, and it is also distinct from fixing the lapse alone as in the previous example.

For example, take $R=\pi_a$ and impose the frame condition $\pi_a=u$. Using
\begin{equation}
\pi_a=-\frac{3a\,\dot a}{4\pi GN},
\end{equation}
we solve for the lapse and substitute it into the action. The resulting dressed Lagrangian is
\begin{equation}
\label{eq:FLRW_pia_dr_action}
\mathcal R_{\pi_a}[L]
=L|_{\pi_a=u}
=\frac{1}{2} u\dot a_\dr-\frac{2\pi G\,u\,a^2\dot\Phi_\dr^2}{3\dot a_\dr}.
\end{equation}

Varying with respect to $a_\dr$ and $\Phi_\dr$ yields \footnote{In the following equations we suppress the $\dr$ labels to keep the expressions readable; all fields and their derivatives in this paragraph are nevertheless dressed fields, and the variations are taken at fixed relational coordinate $u$.}
\begin{subequations}
\begin{align}
\cE_a^\dr
&=-\frac{-8\pi G\,u a^2\ddot a\,\dot\Phi^2
+16\pi G\,u a\,\dot a^2\dot\Phi^2
+4\pi G\,a^2\dot a\dot\Phi(2u\ddot\Phi+\dot\Phi)
+3\dot a^3}{6\dot a^3},\\
\cE_\Phi^\dr
&=\frac{4}{3}\pi G\,a\left[
\frac{a}{\dot a^2}
\bigl(\dot a(u\ddot\Phi+\dot\Phi)-u\ddot a\,\dot\Phi\bigr)
+2u\dot\Phi
\right].
\end{align}
\end{subequations}

The dressed equations are not obtained simply by substituting
the frame condition into the original Euler--Lagrange expressions:
the variation must account for the dependence of the lapse on
the fields introduced when solving $\pi_a=u$. Equivalently,
one can first express the original variation in terms of the
independent equations associated with $a,\Phi,\pi_a$, and then
pull back the resulting equations to the gauge slice.

The solutions of the dressed variational problem are
\begin{equation}
\label{eq:FLRW_pia_dr_solutions}
a\approx\frac{a_0}{u},\qquad
\Phi\approx\Phi_0\pm\sqrt{\frac{3}{4\pi G}}\log u,
\end{equation}
which agree with the original solutions in the gauge $\pi_a=u$.
Thus, on this branch, the dressed variational principle reproduces
the corresponding gauge-fixed solution space.

The contrast with $R=\pi_\Phi$ is instructive. The scalar momentum
\begin{equation}
\pi_\Phi=\frac{a^3\dot\Phi}{N}
\end{equation}
is conserved on shell by $\cE_\Phi=0$. Hence the condition
$\pi_\Phi=u$ is incompatible with the original solution space: 
it attempts to use a constant of motion as a time coordinate.
The associated gauge condition is not transversal to the solution fibers, so it does not define a valid frame there.

Constructing a reduced Lagrangian by solving this condition cannot yield a variational principle equivalent to the original
dynamics; in this model, the resulting equations have no solutions. Explicitly, we have 
\begin{equation}
\label{eq:FLRW_pi_phi_dr_action}
\mathcal R_{\pi_\Phi}[L]
=\frac{1}{2} u\dot \Phi_\dr-\frac{3 u \dot a_\dr^2}{8\pi G a_\dr^2 \dot \Phi_\dr}.
\end{equation}
whose equations of motion are 
\begin{subequations}
\begin{align}
\cE_a^\dr
&=\frac{3 a \left(u \ddot a \dot \Phi +\dot a \left(\dot \Phi -u \ddot \phi \right)\right)-3 u \dot a^2 \dot \Phi}{4 \pi G a \dot \Phi ^2}\\
\cE_\Phi^\dr
&=\frac{6 u \dot a^3 \dot \Phi -3 a \dot a \left(2 u \ddot a \dot \Phi +\dot a \left(\dot \Phi -2 u \ddot \Phi \right)\right)-4 \pi  G a^3 \dot \Phi^3}{8 \pi  G a^3 \dot \Phi^3}
\end{align}
\label{eq:pi_phi_eoms}
\end{subequations}
that have no solution (cf App.~\ref{app:no_sol})

This example makes the distinction precise: a candidate frame may transform covariantly and still fail as a relational clock because it is degenerate on shell.

Taken together, these examples show why the admissibility of a frame and the effect of dressing must be assessed at the level of
the variational principle, not from transformation properties alone. The scalar field and $\pi_a$ provide admissible frames on the branches considered: their dressed variational principles retain the independent equations, with the remaining equation recovered through the Noether identity. The lapse condition $N=1$, by contrast, is a proper-time gauge choice rather than a
relational dressing; imposed before variation, it removes the lapse equation unless the constraint is imposed separately.
Finally, $\pi_\Phi=u$ fails as a frame because $\pi_\Phi$ is constant on shell.

The relevant criterion for a relational frame is therefore twofold. First, it must define a non-degenerate coordinate along the gauge fibers on the solution space of interest. Second, the dressed variational principle must retain enough independent equations to recover the original constrained dynamics. The first condition ensures that the relational description is well defined; the second ensures that its variational principle describes the same classical solution space.

\section{Covariance of the dressed theory and relational Hamiltonians}
\label{sec:cov_rel_Ham}

The following section provides a concrete realisation of the more general formalism discussed in \cite{AFHMtime}, regarding the absence of general covariance in the dressed theory and the emergence of the relational time evolution as its consequence.
However, it is also worth stressing that this discussion differs from the analysis of boundary symmetries in \cite{Carrozza:2022xut}: the present minisuperspace model has
no spatial boundary, and hence no boundary-supported
diffeomorphisms or associated corner charges. Nevertheless,
it admits an integrable bulk generator for evolution in relational
time. This is the classical covariant counterpart of the Page--Wootters
Hamiltonian. Integrability alone does not make it a symmetry
generator: when the dressed Lagrangian depends explicitly on the
relational time, the generator is generally not conserved.

We focus first on the choice of frame $R=a$. It provides a useful
example of how explicit dependence on relational time affects the
symmetries and the conservation of the relational Hamiltonian.
We defer to appendix \ref{app:scalar} for the complementary discussion of the choice of $\Phi$ as a clock.

By pulling back the spacetime metric, we obtain in relational spacetime
\begin{equation}
\d s^2=-N_\dr^2\d u^2+a_\dr^2\d\mathbf{x}^2,
\end{equation}
where $u$ is now the relational time coordinate.

For a transformation generated by the vector field $\rho=\rho(u)\partial_u$ to be a geometric transformation, the dressed metric and scalar field transform as
\begin{equation}
\delta_\rho a_\dr=\rho\,\partial_u a_\dr,\qquad
\delta_\rho N_\dr=\partial_u(\rho N_\dr),\qquad
\delta_\rho\Phi_\dr=\rho\,\partial_u\Phi_\dr.
\label{eq:FLRW_rel_space_diff}
\end{equation}

In the case in which the dressing time is $R=a$, by construction
$a_\dr=u$, which also implies $\partial_ua_\dr=1$.
This last condition is the \textit{dressing condition}: it specifies which
field is to be used as the frame, but does not yet fix the
orientation of the frame.

The first equation in \eqref{eq:FLRW_rel_space_diff} therefore
does not represent a transformation of a dynamical field:
$a_\dr=u$ is no longer varied in the dressed variational principle, and its transformation is instead the transformation of the background relational coordinate.
Although this is consistent with the covariance of the original
bare theory, the roles of $a_\dr$, $N_\dr$ and $\Phi_\dr$ as
background structure or dynamical fields have changed.

This is the origin of the distinction between covariance of
the dressing prescription and covariance of the standalone
relational theory. The field-dependent map defining the dressing
is geometric, and its pullback preserves the tensorial character
of the original fields. However, after the frame has been promoted
to a fixed coordinate, the relational variational principle varies
only the remaining dynamical fields \cite{AFHMtime}.
Consequently, the induced transformation on the relational dynamical fields
alone is not, in general, a symmetry of the relational action.

Explicitly, we have
\begin{equation}
\label{eq:FLRW_dressed_a_Lagr}
L^\dr_a:=\mathcal R_a[L]
=\left(
\frac{u^3\dot\Phi_\dr^2}{2N_\dr}
-\frac{3u}{8\pi GN_\dr}
\right)\d u,
\end{equation}
where $u=a^\dr$ and $\mathcal L^\dr$ denotes the Lagrangian
density.

We now ask whether the transformation
\eqref{eq:FLRW_rel_space_diff} is a symmetry of the dressed
theory. The relevant variation is the one acting only on the
dynamical fields in relational spacetime, which we denote by
$\bar\delta$. In particular, $\bar\delta a_\dr=0$, while $\bar\delta N_\dr\neq 0 \neq \bar\delta \Phi_\dr$.

Using the notation $\check\rho$ for the field-space vector field
induced by $\rho$, we obtain
\begin{equation}
\check\rho\cdot\bar\delta L_\dr
=
-\rho\,\d_u(N_\dr\cE^N_\dr)\d u
+\dot\rho
\underbrace{\left(
\frac{u^3\dot\Phi_\dr^2}{2N_\dr}
+\frac{3u}{8\pi GN_\dr}
\right)}_{\not\approx\,0}
\d t
-\rho\,\partial_u^{\rm exp}\mathcal L^\dr\,\d u.
\label{eq:FLRW_relational_symmetries}
\end{equation}

Here $\partial_u^{\rm exp}$ denotes the explicit time derivative, holding the dynamical fields fixed. The expression is not, in
general, a total derivative. Thus arbitrary relational diffeomorphisms are not symmetries of the dressed theory.

There is an important qualification concerning constant
translations. For $\rho=\mathrm{const.}$, the induced transformation
of the dynamical fields coincides with their time evolution:
\begin{equation}
\delta_\rho\Phi_\dr=\rho\,\dot\Phi_\dr,\qquad
\delta_\rho N_\dr=\rho\,\dot N_\dr.
\end{equation}
A constant translation is a symmetry only if the relational
Lagrangian is autonomous, i.e. if it has no explicit dependence
on $u$. For the $a$-frame, the explicit factors of $u$ prevent
this. Thus even relational time translations are not symmetries
in this example.

We emphasize that, because we are varying only the relational
dynamical fields, the action of $\check\rho\cdot\bar\delta$ on
the dressed fields is not generally a geometric transformation (in the sense of being the image under $\cR$ of the action  of a vector field action from spacetime to relational space):
\begin{equation}
\check \rho \cdot \bar \delta \psi^\dr
\neq\pounds_\rho\psi^\dr = \cR\left[\pounds_{\cR^-1[\rho]} \psi\right].
\end{equation}
It acts geometrically on the dynamical components of the dressed
metric, but not on the background structure $a^\dr=u$.\\

The theory described by the dressed Lagrangian
\eqref{eq:FLRW_dressed_a_Lagr} is a simple mechanical model and
can be treated with the usual machinery of classical mechanics.
It is also convenient to use a more covariant notation, which
will be useful when comparing the dressed theory with its
canonical formulation.

Because $a$ is a good frame, the equations of motion obtained
from \eqref{eq:FLRW_dressed_a_Lagr} form a system of independent
ordinary differential equations. Its general solution (in the patch of relational space where $u>0$) is
\begin{equation}
\label{eq:FLRW_a_sol}
\Phi^\dr\approx\Phi_0
\pm\sqrt{\frac{3}{4\pi G}}\log(u),
\qquad
N_\dr=N_0\, u^2.
\end{equation}
These are precisely the solutions of
\eqref{eq:FLRW_eoms} in the gauge $a=u$, for $N_0 := 3/a_0^3$.\\

\paragraph{The dressed symplectic structure.}

The presymplectic structure can be constructed in two related ways, which differ in how they treat variations of the relational frame. The distinction is important off-shell, although the two constructions yield the same physical symplectic structure once the constraints are imposed.

The most straightforward procedure is to evaluate the symplectic current from the variational principle of the dressed action
\eqref{eq:FLRW_dressed_a_Lagr}:
\begin{equation}
\delta L^\dr
=
\left(
\cE_\Phi^\dr\delta\Phi^\dr
+\cE^N_\dr\delta N_\dr
+\d_u(\pi_\Phi^\dr\delta\Phi_\dr)
\right)\d u,
\qquad
\pi_\Phi^\dr=\frac{u^3}{N_\dr}\dot\Phi^\dr.
\end{equation}

This procedure varies the dressed fields at fixed relational coordinate $u$. In particular, it does not include an independent variation of the frame map used to define that coordinate. To keep track of such variations, one can instead pull back the presymplectic current of the original theory along the field-dependent frame map.

As shown in App.~\ref{app:covariant phase space}, the pullback is naturally expressed in terms of the dressed presymplectic current
\begin{equation}
\omega_\varpi
=
\omega-\iota_\varpi{\rm Eom},
\end{equation}
up to a spacetime-exact contribution. Here $\varpi=\delta R^{-1}\circ R$ is the field-space one-form associated with variations of the inverse frame map, and the spacetime contraction
$\iota_\varpi{\rm Eom}$ accounts for the fact that the relational
coordinates themselves depend on the fields.

The current $\omega_\varpi$ is still expressed as a field-space
two-form evaluated at a spacetime point of the original manifold.
Consequently, its evaluation at a fixed coordinate time $t$ is
not, in general, independent of that choice (off-shell). On shell,
the conservation of the presymplectic current makes it
independent of the choice of time. The pullback to relational spacetime
resolves this ambiguity even off-shell by specifying the evaluation point through
the frame: the resulting current is evaluated at fixed relational
coordinate $u$, rather than at a coordinate time whose
identification is itself gauge dependent.

In a theory with a spatial boundary, the spacetime-exact
contribution can give a corner term and must be treated together
with the boundary conditions. In the present minisuperspace model
there is no spatial boundary, so this corner contribution is absent. Possible contributions at temporal endpoints are a separate issue
and depend on the variational principle and endpoint conditions.

For the choice $R=a$, the resulting current is
\begin{align}
\omega_\varpi
&=\omega-\iota_\varpi{\rm Eom}
\nonumber\\
&=\delta\pi_\Phi\curlywedge\delta\Phi
+\delta\,\pi_a\curlywedge\delta a
+\frac{\delta a}{\dot a}\curlywedge
(\cE_\Phi\delta\Phi+\cE_N\delta N)
\nonumber\\
&=
\left(\delta\pi_\Phi
-\frac{\delta a}{\dot a}\dot\pi_\Phi\right)
\curlywedge
\left(\delta\Phi
-\frac{\delta a}{\dot a}\dot\Phi\right)
+\frac{\delta a}{\dot a}\curlywedge
\delta(N\cE_N).
\end{align}

Pulled back to relational spacetime, this becomes
\begin{equation}
\mathcal R_a[\omega_\varpi]
=
\delta\pi_\Phi^\dr\curlywedge\delta\Phi_\dr
+
\mathcal R_a\left[\frac{\delta a}{\dot a}\right]
\curlywedge\delta(N_\dr\cE^N_\dr).
\label{eq:Omega_dress}
\end{equation}

The significance of this pullback is that the resulting expression
is evaluated at the point where $a=u$. The relational coordinate
$u$ is a label on the relational spacetime: it is unchanged under
a simultaneous diffeomorphism of the fields and the frame map.
Thus, unlike the original current evaluated at a prescribed
coordinate time $t$, the pulled-back current specifies its
spacetime evaluation relationally, including away from the
equations of motion.

In the language of \cite{Carrozza:2022xut}, the first term corresponds to the \textit{invariant} part of the symplectic structure, while the second term is the \textit{gauge part}.
The second term contains the variation of the frame coordinate
and is proportional to the dressed constraint. It is therefore absent when the constraint is imposed and the
presymplectic structure is pulled back to the constraint
surface, so that allowed tangent variations satisfy
$\delta(N_\dr\cE^N_\dr)=0$. On this restricted space the
second term vanishes, and the induced symplectic structure is
\begin{equation}
\label{eq:Omega_phys}
\Omega^{\rm phys}
=
\delta\pi_\Phi^\dr\curlywedge\delta\Phi_\dr .
\end{equation}
Notice that $\pi_\Phi^\dr$ depends explicitly on time at the
covariant level, but is an independent phase-space variable at
the canonical level. On shell, we also have
$\d_u\pi_\Phi^\dr=0$. Also, using the general solutions \eqref{eq:FLRW_general_solutions}, the symplectic structure becomes conserved in time, and hence\footnote{The presence of the $\pm$ indicates the presence of the two branches also at the level of the physical phase space. There is however a degeneracy coming from the possibility of changing branches either by picking a different sign in the overall $\pm$, or by flipping the signs of the parameters $\Phi_0$, $a_0$}
\be
\Omega^{\rm phys} \approx  \omega \approx \pm \f{1}{\sqrt{12 \pi G}} \delta a_0^3 \curlywedge \delta \Phi_0 
\ee

The comparison between the two constructions can now be made
precise. The variation of the standalone dressed action is
performed at fixed relational coordinate and yields the
presymplectic structure associated with the dressed configuration
variables. By contrast, pulling back the original presymplectic
current also accounts for the field dependence of the frame map.
It therefore retains an additional term, conjugate to variations
of the frame coordinate and proportional to the dressed
gauge constraint. The two constructions need not agree
off shell, but their difference vanishes upon restriction to the
physical constraint surface. In this sense, the additional term
records the frame direction and its associated constraint; it
does not represent an additional physical degree of freedom.

\subsection{Relational phase space and Hamiltonian}

The absence of relational time-translation invariance does not
prevent us from defining a Hamiltonian that generates evolution
in relational time. Rather, it means that this generator need not
be conserved. The canonical phase space is obtained through the
Legendre transform of the dressed action, which, in the present
model, involves only $\dot\Phi_\dr$, the sole dynamical variable
with a kinetic term. We first distinguish the covariant momentum
from its canonical phase-space representative. The dressed
covariant momentum is the field-dependent quantity
\begin{equation}
\pi_\Phi^\dr
:=\frac{\partial\cL^\dr}{\partial\dot\Phi_\dr}
=\frac{u^3\dot\Phi_\dr}{N_\dr}.
\end{equation}
By contrast, $P_\Phi$ denotes the canonical momentum regarded as
an independent coordinate on relational phase space. The
Legendre map identifies the two,
$P_\Phi=\pi_\Phi^\dr$, but their roles are different: the
expression for $\pi_\Phi^\dr$ depends explicitly on $u$ (not only through
the dressed fields, but explicitly on the coordinate $u$), whereas $P_\Phi$ is a phase-space variable
without explicit time dependence, whose evolution is determined by Hamilton’s equations. The Legendre transform gives
\begin{align}
H^\dr
&=\pi_\Phi^\dr\dot\Phi_\dr-\cL^\dr
\nonumber\\
&=
\frac{u^3(\dot\Phi_\dr)^2}{2N_\dr}
+\frac{3u}{8\pi GN_\dr}
\nonumber\\
&=
\frac{N_\dr}{2u^3}(P_\Phi)^2
+\frac{3u}{8\pi GN_\dr}.
\label{eq:H_rel}
\end{align}
In the last line we have expressed the Hamiltonian in terms of
the canonical momentum, as usual in the Legendre transform.
The Poisson brackets below are those of the relational
canonical phase space, with $(\Phi_\dr,P_\Phi)$ as a canonical
pair.

The Hamiltonian generates the relational evolution equations
on the physical phase space \eqref{eq:Omega_phys}:
\begin{equation}
\begin{aligned}
\dot P_\Phi
&=\{P_\Phi,H^\dr\}=0,\,,
\qquad
\dot\Phi_\dr
&=\{\Phi_\dr,H^\dr\}
=\frac{N_\dr}{u^3}P_\Phi\,.
\end{aligned}
\end{equation}
Using the Legendre relation $P_\Phi=\pi_\Phi^\dr$ (which trivially solves the second equation), these equations are equivalent to
\begin{equation}
\dot P_\Phi=0
\quad\Longleftrightarrow\quad
\d_u\pi_\Phi^\dr=0\,.
\end{equation}

The presence of a constraint in the dressed Hamiltonian does
not, by itself, imply a residual gauge redundancy on relational
spacetime. To see this, recall that $N_\dr$ has no conjugate
velocity and therefore acts as a Lagrange multiplier. Its
conjugate momentum gives the primary constraint
\begin{equation}
P_N\approx0.
\end{equation}
Preservation of this constraint under the total Hamiltonian
$H_T=H^\dr+\lambda P_N$ yields a secondary constraint:
\begin{equation}
\begin{aligned}
0\approx\{P_N,H_T\}
&=-\frac{\delta H^\dr}{\delta N_\dr}
\\
&=-\frac{1}{2u^3}P_\Phi^2
+\frac{3u}{8\pi G(N_\dr)^2}
=:\cH^\dr.
\end{aligned}
\end{equation}
Under the Legendre relation $P_\Phi=\pi_\Phi^\dr$, this
secondary constraint coincides with the dressed lapse equation,
up to a sign convention
\begin{equation}
\cH^\dr\approx-\cE^N_\dr.
\end{equation}

The primary and secondary constraints do not generate residual
gauge transformations. Indeed, their Poisson bracket is
\begin{equation}
\{P_N,\cH^\dr\}
=
\frac{3u}{4\pi G(N_\dr)^3}\neq0.
\end{equation}
They therefore form a second-class pair on the domain where
$u\neq0$ and $N_\dr\neq0$. In particular, the secondary
constraint fixes the dressed lapse rather than generating an
arbitrary transformation of the relational fields. This is
analogous to the usual conversion of first-class constraints
into second-class constraints by gauge-fixing, although here
the second-class structure arises directly in the relational
description. The time evolution of the dressed lapse is consequently fixed.
Indeed,
\begin{equation}
\dot N_\dr\approx\{N_\dr,H_T\}=\lambda,
\end{equation}
while conservation of the secondary constraint determines
$\lambda$:
\begin{align}
0&\approx
\partial_u\cH^\dr+\{\cH^\dr,H_T\}
\nonumber\\
&=\partial_u\cH^\dr
+\cancel{\{\cH^\dr,H^\dr\}}
+\lambda\{\cH^\dr,P_N\}.
\end{align}
Hence
\begin{equation}
\dot N_\dr
\approx
-\frac{\partial_u\cH^\dr}{\{\cH^\dr,P_N\}}
=
\frac{N_\dr\left(
4\pi G\, P_\Phi^2 \,N_\dr^2+u^4
\right)}{2u^5},
\end{equation}
which agrees with the solutions \eqref{eq:FLRW_a_sol}.

The dressed lapse thus remains a Lagrange multiplier, but its time dependence is fixed by the consistency conditions of the second-class constraints, leaving no residual freedom to prescribe it arbitrarily. This illustrates why the existence of constraints must be distinguished from the presence of gauge redundancy: the latter is associated with first-class constraints, whereas the dressed primary and secondary constraints restrict the allowed relational configurations and determine the lapse.

The extended canonical pair $(P_N,N)$ plays a role analogous to the additional term in \eqref{eq:Omega_dress}: both retain extra directions in the phase space that encode aspects of the covariance of the undressed theory. These directions do not represent a gauge redundancy in the dressed description; rather, they are eliminated by the corresponding equations of motion and constraints. There is, however, a difference between the two constructions: the dressed presymplectic structure \eqref{eq:Omega_dress} already incorporates the secondary constraint, whereas the extended canonical formulation also includes the primary constraint $P_N\approx0$ associated with the lapse momentum. \\

We now turn to the covariant characterization of the generator
of relational evolution. This also clarifies its relation to
time-translation invariance and to Noether charges. Consider
the transformation induced by a constant translation of
relational time. Its action on the dressed fields coincides
with their evolution in $u$. The corresponding generator can
be integrable even when the dressed Lagrangian is not invariant
under this transformation. In that case, it generates
relational evolution but is not a conserved Noether charge.

For constant relational time translations, the non-covariance
of the Lagrangian \eqref{eq:FLRW_relational_symmetries} is
directly related to its explicit dependence on relational time.
In terms of the relational anomaly operator introduced in
App.~\ref{app:covariant phase space},
\begin{equation}
\label{eq:relational_lagrangian_anomaly}
\partial_u\rho=1
\quad\Longrightarrow\quad
0=
\rho\,\partial_u\cL^\dr\,\d u
+\check\rho\cdot\bar\delta L
-\d\iota_\rho L
:=
\rho\,\partial_u\cL^\dr\,\d u
+\bar\Delta_\rho L,
\end{equation}
where $\bar\Delta_\rho$ is defined in
\eqref{eq:anomaly operator}~\cite{Carrozza:2022xut} in terms of spacetime covariance, here lifted to the relational spacetime.
This identity separates the explicit relational-time dependence
of the Lagrangian from its variation under the induced
transformation of the fields. In particular, the first term
is the obstruction to invariance under relational time
translations.

Similarly, for any $\rho$, possibly field dependent, we have
(here $\theta^\dr$ is a zero-form in spacetime):
\begin{equation}
\label{eq:non_cov_theta_FLRW}
\rho\,\partial_u^{\rm exp}\theta^\dr
+\bar\Delta_\rho\theta^\dr
=
\rho\,\partial_u^{\rm exp}\theta^\dr
+\check\rho\cdot\bar\delta\theta^\dr
+\bar\delta(\check\rho\cdot\theta^\dr)
-\iota_\rho\d\theta^\dr
-\widecheck{\delta\rho}\cdot\theta^\dr
=0.
\end{equation}
The explicit derivative $\partial_u^{\rm exp}$ acts only on
the relational-time dependence of the covariant quantity,
holding the dynamical fields fixed. In particular, it captures
the explicit time dependence of the symplectic potential,
which in this model is encoded in $\pi_\Phi^\dr$.

For a constant unit translation, the covariant expression
$H_\rho=\check\rho\cdot\theta^\dr-\iota_\rho L^\dr$
reduces, with the conventions used here, to the Legendre
Hamiltonian $H^\dr$ of the standalone dressed mechanical system \eqref{eq:H_rel}. The non-covariance of the Lagrangian then leads to the time evolution of the relational  generator:
\begin{equation}
\d H_\rho
=
\d \bigl(
\check\rho\cdot\theta^\dr-\iota_\rho L^\dr
\bigr)
=
-\check\rho\cdot{\rm Eom}^\dr
-\rho\,\partial_u\cL^\dr\,\d u.
\end{equation}
On shell,
\begin{equation}
\d H_\rho
\approx
-\rho\,\partial_u\cL^\dr\,\d u.
\end{equation}
Consequently, the generator is not conserved in general:
its variation is controlled by the explicit relational-time
dependence of the Lagrangian. If the latter vanishes, the
generator is conserved on shell.
This is the situation, for example, in the $\Phi$-frame
considered in the appendix \ref{app:scalar}.

Using the corresponding relation for the symplectic potential, now for a general $\rho$, we obtain
\begin{equation}
\check\rho\cdot\omega^\dr
+\rho\,\partial_u\theta^\dr
=
-\delta H_\rho-\iota_\rho{\rm Eom}^\dr.
\end{equation}
Thus the flow generated by $\rho$ always admits an integrable
on-shell generator. In particular, for $\rho =const$, this relation identifies $H^\dr$ as obtained by a Legendre transform, with $H_\rho$ as the integrable covariant generator associated with a relational transformation. We stress again that it generates relational-time reparametrization; it is not necessarily conserved, it does not, in general,
generate a symmetry of the dressed action, and it does not generate a gauge redundancy, due to the lack of general covariance of the dressed action.

The term $\rho\,\partial_u\theta^\dr$ is essential because
it accounts for the explicit time dependence of the
covariant symplectic potential. In particular,
\begin{equation}
\check\rho\cdot\delta\pi_\Phi^\dr
+\rho\,\partial_u\pi_\Phi^\dr
=
\rho\,\d_u\pi_\Phi^\dr.
\end{equation}

On shell of the secondary constraint
$\cE^N_\dr\approx0$, we have
\begin{equation}
H^\dr
\approx
\frac{3u}{4\pi GN_\dr},
\label{eq:half_onshell_H_FLRW}
\end{equation}
and fully on shell,
\begin{equation}
H^\dr
\approx
\frac{3}{4\pi G\,N_0\,u}\neq0.
\label{eq:onshell_H_FLRW}
\end{equation}
The relational Hamiltonian is therefore non-vanishing and,
for this solution, explicitly dependent on relational time.
It is not a conserved quantity.

The dressed theory consequently has three characteristic
properties:
\begin{itemize}
 \item Generic reparametrizations of relational
time are not symmetries of the dressed action.\footnote{
There is, however, a conformal invariance of the dressed action
\eqref{eq:FLRW_a_sol} under
$u \to \alpha u$, $\Phi_\dr \to \Phi_\dr$,
$N_\dr \to \alpha^2 N_\dr$, for constant parameter $\alpha$. A discussion of the associated Noether charge and Hamiltonian flow goes beyond the scope of this paper, which focuses on relational time evolution. We
therefore leave for future work the possible interplay with Schrödinger-type symmetries
\cite{BenAchour:2022fif,Geiller:2022baq}, usually studied in proper-time or conformal gauges, but potentially realized here directly in relational variables.
}
Constant translations of relational time are
symmetries only when the dressed Lagrangian is independent of
$u$.

\item An integrable generator of relational evolution can be
defined even in the absence of such symmetries. Unless the
Lagrangian is autonomous, this generator is not conserved.
Thus, the existence of a Hamiltonian generating
relational evolution does not require the existence of a
corresponding time-translation symmetry.

\item The dressed theory retains constraints inherited from
the original gauge constraints. These form a second-class
pair and therefore do not generate residual gauge
redundancies. Instead, they restrict the relational
configurations and fix the time dependence of the dressed
lapse.
\end{itemize}

These properties lead to an important distinction between
relational evolution and gauge evolution, also illustrated in Figures~\ref{fig:gauge_section} and~\ref{fig:dressed_field}. Although the dressed
action still yields a constraint through variation of the
lapse, this does not reintroduce a gauge redundancy on
relational spacetime. The constraints instead restrict the
relational phase space, while the relational equations define
an independent system of ordinary differential equations.
Their solutions are determined by a finite number of initial
data, and relational evolution is generated by a Hamiltonian
flow on the reduced relational phase space. This flow need
not arise from a symmetry of the dressed action, and its
generator need not be conserved.

Finally, the relational Hamiltonian considered here is a bulk
quantity: it arises from the Legendre transform of the
relational action, not from a boundary contribution, consistently with the discussion in \cite{AFHMtime,JTproj}. It is
therefore distinct in origin from charges associated with
boundary-supported transformations, as discussed in
\cite{Carrozza:2022xut}. In theories with spatial boundaries,
the relational description may involve additional boundary
data and associated charges. These features are absent in the
homogeneous minisuperspace model considered here.

\section{Canonical relational dynamics and gauge-fixing}
\label{sec:canonical}

We now compare the covariant dressing with the canonical
description of the FLRW system, keeping in view its relation
to Page--Wootters evolution. The analysis below is entirely
classical: it concerns complete observables and their
Hamiltonian generators, not a construction of conditional
quantum states. We first construct the classical
canonical relational observables associated with a dynamical frame on the phase space and
identify the corresponding generator of reorientations. We then compare this construction with canonical
gauge-fixing and reduction, in order to clarify how the
covariant and canonical relational Hamiltonians are related.

We begin with the kinematical phase space in spacetime, as in the usual ADM description,
\begin{equation}
\omega^{\rm kin}
=
\delta\,p_a\curlywedge\delta a
+\delta p_\Phi\curlywedge\delta\Phi.
\end{equation}
Here $p_a$ and $p_\Phi$ are independent canonical momenta,
rather than functions of the configuration variables. On the
Legendre image of the original Lagrangian, they coincide with
the corresponding covariant momenta $\pi_a$ and $\pi_\Phi$.
The Hamiltonian constraint is
\begin{equation}
\cH
=
-\cE_N
=
\frac{1}{2a^3}p_\Phi^2
-\frac{2\pi G}{3a}p_a^2.
\label{eq:FLRW:constr}
\end{equation}
For clarity, we distinguish throughout between the canonical coordinates $(a,\Phi,p_a,p_\Phi)$, the velocity-dependent covariant momentum functions $(\pi_a,\pi_\Phi)$ defined by the Lagrangian presymplectic potential, the relational canonical momentum $P_\Phi$, the gauge-fixed lapse $N_{g.f.}$, and the dressed lapse $N_\dr$. The symbols $p_a,p_\Phi$ denote independent phase-space coordinates; they coincide with $\pi_a,\pi_\Phi$ only on the Legendre image of the original Lagrangian.\\

\paragraph{Canonical relational observables.}

For a dynamical frame $R$, the classical counterpart of the
quantum dressing can be expressed in terms of complete
observables \cite{Dittrich:2004cb,Dittrich:2005kc,Tambornino:2011vg}.
These observables describe the value of a phase-space function
when the frame variable takes a prescribed value $R_0$.
In this sense, they implement a classical conditional description:
the system observable is evaluated relative to a specified
reading of the clock, rather than at a prescribed coordinate
time. This is the classical analogue of the relational evolution
in the Page--Wootters construction, where the state of a system
is conditioned on a reading of a quantum clock. The comparison
is nevertheless limited to the classical level. We do not
construct a physical Hilbert space, define conditional quantum
states or operators, or address the clock-system factorization
and quantum correlations required for the Page--Wootters
formalism. Our aim here is instead to compare the classical
relational dynamics obtained from covariant dressing with
that generated by canonical complete observables.

The complete observable associated with a phase-space function
$g$ and the frame value $R_0$ is
\begin{equation}
\label{eq:canonical_complete_observable}
\cO_g^{R_0}
=
\int\d t\,
\delta(R_0-\alpha_t(R))\,
\alpha_t(\{R,\cH\})\,
\alpha_t(g),
\end{equation}
where, for any phase-space function $g$,
\begin{equation}
\alpha_t(g)=\exp(t\{\cdot,\cH\})\,g.
\end{equation}
This expression evaluates $g$ at the point along the
Hamiltonian-constraint orbit where the frame takes the value
$R_0$. We assume that the frame intersects the relevant orbit
transversally and that the intersection is unique in the
region under consideration. The factor $\alpha_t(\{R,\cH\})$
accounts for the Jacobian associated with this intersection;
the formula is understood with the orientation of the orbit
fixed on the branch being considered.

At the classical level, the construction of relational
observables is insensitive to a rescaling of the constraint
by a phase-space function, provided the rescaling is
non-vanishing and regular on the relevant region of the
constraint surface. In particular, consider
\begin{equation}
X=f\cH.
\end{equation}
On shell,
\begin{equation}
\tilde\alpha_t(g)
:=
\exp(t\{\cdot,f\cH\})\,g
\end{equation}
satisfies
\begin{equation}
\partial_t\tilde\alpha_t(g)
=
\{\tilde\alpha_t(g),f\cH\}
\approx
f\{\tilde\alpha_t(g),\cH\}.
\end{equation}
Consequently, the two flows agree up to a reparametrization
of their orbit parameter:
\begin{equation}
\tilde\alpha_t(g)\approx\alpha_\tau(g),
\qquad
\frac{\d\tau}{\d t}=f.
\end{equation}
Thus, as long as $f$ is non-vanishing and regular, rescaling
the constraint changes the parametrization of its orbits
without changing the orbits themselves. The corresponding
relational observables therefore agree:
\begin{align}
\widetilde{\cO}_g^{R_0}
&=
\int\d t\,
\delta(R_0-\tilde\alpha_t(R))\,
\tilde\alpha_t(\{R,f\cH\})\,
\tilde\alpha_t(g)
\nonumber\\
&\approx
\int\d t\,
\delta(R_0-\alpha_\tau(R))\,
f\alpha_\tau(\{R,\cH\})\,
\alpha_\tau(g)
\nonumber\\
&=\cO_g^{R_0}.
\end{align}
In the last step, we have changed the integration variable from $t$ to $\tau$, with $\d\tau=f\d t$.

This reparametrization is valid only on regions of phase space where $f$ is non-vanishing and sufficiently regular. If $f$ vanishes on the constraint surface, the redefinition can cease to be invertible, and the flow generated by $f\cH$ may fail to traverse the same gauge orbit. Such points must be treated separately: they can split the domain into branches or alter the global definition of the relational observable. The equivalence asserted here is therefore local to a branch on which the rescaling is admissible. In addition, if $f<0$, the orbit parameter is orientation-reversed; the change of integration variable in the complete-observable formula must then use the correspondingly oriented integration domain. The equality of the two expressions assumes that this orientation and the range of integration are transformed consistently. It is convenient to choose $f=a^3$. The rescaled constraint is then
\begin{equation}
X
=
\frac12 p_\Phi^2
-\frac{2\pi G}{3}a^2p_a^2
\approx0.
\end{equation}
Now redefining the flow as $\alpha_t(g):=\exp(t\{\cdot,X\})g$, we find
\begin{equation}
\alpha_t(a\,p_a)=a\,p_a,
\end{equation}
and
\begin{equation}
\alpha_t(a)
=
a\exp\left(-t\frac{4\pi G\,a\,p_a}{3}\right),
\qquad
\alpha_t(p_a)
=
p_a\exp\left(t\frac{4\pi G\,a\,p_a}{3}\right).
\end{equation}

The generator of relational evolution associated with the $a$-frame can be obtained canonically by constructing the complete observable of the momentum conjugate to the frame.
\be
\cO_{p_a}^{a=u} &= \int \d t \,\delta \left (u -\alpha_t(a)\right) \alpha_t(p_a)  \alpha_t \left(\{a, X\}\right)\notag \\ 
&=- \int \d t\, \delta \left (u -a \exp \left(-t \f{4\pi G\, a\,p_a}{3}\right) \right) p_a  \exp \left(t \f{4\pi G\, a\,p_a}{3}\right) \f{4 \pi G}{3}  a^2 p_a \exp \left(-t \f{4\pi G\, a\,p_a}{3}\right)\notag \\ 
&= -\f{1}{u} a\, p_a
\ee
This is the complete observable associated with $p_a$,
evaluated when the frame takes the relational value $a=u$.
On the corresponding gauge section, $a=u$, it therefore
reduces to $-p_a$. The resulting expression is proportional
to the inverse clock reading, as is the covariant relational
Hamiltonian in \eqref{eq:onshell_H_FLRW}. This motivates
identifying the clock reading in the complete observable
with the relational coordinate used in the covariant
description.

To make the correspondence explicit, we use the Legendre relation
\begin{equation}
a\,p_a \to a\,\pi_a=-\frac{3a^2\dot a}{4\pi GN},
\end{equation}
where $\pi_a$ is the momentum function defined by the
covariant Lagrangian. This substitution requires some care: even though $a\,p_a$ commutes with the constraint,
$a\,\pi_a$ transforms as a scalar under time
reparametrizations, its value at a prescribed coordinate time
is not thereby an off-shell diffeomorphism-invariant observable. Similarly, the kinematical presymplectic structure at fixed coordinate time retains an explicit dependence on the choice of spacetime slice. The Legendre relation alone therefore does not turn the canonical complete observable into a covariant relational quantity.

The missing step is to either restrict to the solution space, where $a\,\pi_a$ is a constant in time, or, dressing it again,
thereby specifying its evaluation point through the same
relational frame. Identifying the clock reading in
$\cO_{p_a}^{a=u}$ with the relational coordinate $u$, we obtain
\begin{equation}
-\f{1}{u} \cR\left[ a\, \pi_a  \right] = -\cR\left[\pi_a  \right] = 
\frac{3u}{4\pi GN_\dr}
\approx H^\dr.
\end{equation}
The final identification with $H^\dr$ holds on the constraint
surface. Three objects should therefore be kept distinct:
(i) $\cO_{p_a}^{a=u}$, the canonical complete observable;
(ii) $\cR_a[\pi_a]$, the covariantly dressed
momentum function; and (iii) $H^\dr$,
the Legendre Hamiltonian of the standalone dressed action.
The correspondence between them involves the Legendre map,
relational evaluation of the resulting expression, and the
Hamiltonian constraint (or, equivalently, restriction to the
full solution space). It is not a direct off-shell equality
between the three quantities.

This distinction may matter when going to the quantum theory.  In particular, the order of integration becomes relevant, and the potential contributions from off-shell trajectories become relevant when going beyond saddle point approximations
Whether the two constructions define
equivalent insertions or operators, depends on the
quantization procedure and the treatment of the lapse and
constraint. We leave this question for future work.

\subsection{Gauge-fixing and the reduced phase space}

In many gravitational systems, a convenient relational frame
cannot be identified directly in terms of the fundamental
fields. It is therefore useful to formulate the construction
in canonical gauge-fixing language, where gauge conditions and their
preservation can be treated more systematically \cite{JTproj}.

A dressing condition can be understood as a differential
condition that the dressed fields must satisfy, analogous
to $\partial_\mu A^\mu_\dr=0$ for Lorenz gauge or
$A^\dr_r=0$ for radial Wilson lines. Such a condition
does not, by itself, specify the full dressing: one must
also fix the orientation of the frame, corresponding to
the initial value of the Wilson lines. Without this
additional information, a condition on the dressed fields
need not define a complete gauge-fixed variational problem.

In the canonical formulation, this distinction can be made
more explicit. We start from the extended phase space,
including the lapse and its conjugate momentum, with
first-class constraints $\phi_i\approx0$ and canonical
Hamiltonian $H_{\rm c}$. The corresponding extended
Lagrangian is
\begin{equation}
\mathcal L^{\rm ext}
=
p_a\,\dot a+p_\Phi\dot\Phi+\pi_N\dot N-H_{\rm c},
\end{equation}
where
\begin{equation}
H_{\rm c}:=\mu\pi_N+N\cH,
\qquad
\phi_1=\pi_N,
\qquad
\phi_2=\{\phi_1,H_{\rm c}\}=\cH,
\qquad
\{\phi_1,\phi_2\}=0.
\end{equation}
The integrable generator of gauge transformations is
\begin{equation}
G_\epsilon=\dot\epsilon\pi_N+\epsilon\cH.
\end{equation}
It generates the field-dependent diffeomorphisms
\eqref{eq:FLRW_gauge_transf} with $\xi=\epsilon/N$.

We now impose the condition $\dot a=1$, which fixes the
lapse in terms of the canonical variables. At the canonical
level, this condition reads
\begin{equation}
\{a,H_{\rm c}\}=1
\quad\Longrightarrow\quad
-\frac{4\pi GNp_a}{3a}=1.
\end{equation}
Thus,
\begin{equation}
0=\chi:=N-N_{g.f.}
=N+\frac{3a}{4\pi Gp_a}.
\label{eq:FLRW_gauge_fix}
\end{equation}

This condition is only a partial gauge-fixing, because it
still leaves residual time translations. To see this, we
impose
\begin{equation}
0\overset{!}{=}\{\chi,G_\epsilon\}
=
\dot\epsilon-\epsilon\{N_{g.f.},\cH\}
=
\dot\epsilon-\epsilon\frac{\dot N_{g.f.}}{N}.
\label{eq:FLRW_residual_gauge}
\end{equation}
Here we used the fact that $N_{g.f.}$ is a function of the
original, unextended kinematical phase space, so that
\begin{equation}
\dot N_{g.f.}
=
\{N_{g.f.},N\cH\}
=
N\{N_{g.f.},\cH\}.
\end{equation}

On the gauge-fixing surface, the residual gauge transformations
are therefore parametrized by
\begin{equation}
\epsilon_{\rm res}=\epsilon_0N_{g.f.},
\end{equation}
where $\epsilon_0$ is constant.

It is convenient to reparametrize the original first-class
constraint set in terms of
\begin{equation}
C_\chi=\pi_N,
\qquad
C_{\rm res}=\{N_{g.f.},\cH\}\pi_N+\cH.
\end{equation}
By construction, the gauge-fixing condition makes
$\{\chi,C_\chi\}$ second class, while $C_{\rm res}$ remains
first class:
\begin{equation}
\{\chi,C_\chi\}=1\not\approx0,
\qquad
\{\chi,C_{\rm res}\}
\approx
\{N_{g.f.},\cH\}
-\{N_{g.f.},\cH\}
=0.
\end{equation}
Moreover, the residual gauge generator is
\begin{equation}
G_\epsilon^{\rm res}
=
\epsilon\{N_{g.f.},\cH\}\pi_N+\epsilon\cH
=
\epsilon C_{\rm res}.
\end{equation}
On the gauge-fixing surface, the residual parameter is
$\epsilon=\epsilon_0N_{g.f.}$, so that, after imposing
$\pi_N=0$, the residual generator reduces to
$\epsilon_0N_{g.f.}\cH$. The remaining transformations are
therefore generated by the reduced Hamiltonian
$N_{g.f.}\cH$, with constant parameter $\epsilon_0$.

The gauge-fixing condition $\chi=N-N_{g.f.}$ is transversal
to the gauge orbits, except for transformations satisfying
\eqref{eq:FLRW_residual_gauge}. We can therefore regard
$\chi$ as a gauge-fixing for the primary constraint
$\pi_N\approx0$.

Imposing the second-class pair $\chi=0=C_\chi$ eliminates
the lapse and its conjugate momentum, but it does not yet
complete the physical reduction: the first-class residual
constraint $C_{\rm res}$ remains. The resulting partially
gauge-fixed (lapse-eliminated) phase space is isomorphic, as
a symplectic manifold, to the original unextended kinematical
phase space. This can be seen directly by restricting the
extended symplectic form to the gauge section:
\begin{equation}
\begin{aligned}
\omega^{\rm red}
&=
\omega^{\rm ext}|_{\chi=0=C_\chi}
\\
&=
\delta\,p_a\curlywedge\delta a
+\delta p_\Phi\curlywedge\delta\Phi
+\cancel{\delta\pi_N}\curlywedge\delta N_{g.f.}
\\
&\simeq\omega^{\rm kin}.
\end{aligned}
\end{equation}
The isomorphism concerns the symplectic structure and does
not, by itself, identify the reduced momenta with the
velocity-dependent momentum functions of the original
covariant Lagrangian. In particular, although we retain the
notation $p_a$ and $p_\Phi$ for the reduced momentum
coordinates, they must now be understood as coordinates on
the reduced phase space. The Legendre relations used in the
previous section cannot therefore be applied directly to
reconstruct the original configuration-space description
from these variables. This distinction will be important
when comparing the canonical relational dynamics with the
covariantly dressed formulation below. Equivalently, the Dirac bracket is
\begin{equation}
\{F,G\}_{\rm D}
=
\{F,G\}
+
\frac{1}{\{\chi,C_\chi\}}
\left(
\{F,\chi\}\{C_\chi,G\}
-\{F,C_\chi\}\{\chi,G\}
\right).
\end{equation}
For the generators of $\omega^{\rm kin}$, it reduces to
the standard Poisson bracket.

On partially gauge-fixed phase space, the reduced theory is
\begin{align}
\mathcal L^{\rm red}
&=
\mathcal L^{\rm ext}|_{N=N_{g.f.},\,\pi_N=0}
\nonumber\\
&=
p_a\,\dot a+p_\Phi\dot\Phi-N_{g.f.}\cH
\nonumber\\
&=
p_a\,\dot a+p_\Phi\dot\Phi
-\frac{p_a}{2}
+\frac{3}{8\pi Ga^2p_a}p_\Phi^2.
\end{align}

At this level $\mathcal L^{\rm red}$ does not admit a purely
Lagrangian formulation in terms of configuration variables
alone, because the Legendre transform of $N_{g.f.}\cH$
is degenerate. Indeed, solving
\begin{equation}
\dot a=\{a,N_{g.f.}\cH\},
\qquad
\dot\Phi=\{\Phi,N_{g.f.}\cH\}
\end{equation}
for $p_a$ and $p_\Phi$ gives
\begin{equation}
\dot a
=
\frac{3p_\Phi^2}{8\pi Ga^2p_a^2}
+\frac12,
\qquad
\dot\Phi
=
-\frac{3p_\Phi}{4\pi Ga^2p_a}.
\end{equation}
The two momenta only appear in the relative combination
$p_\Phi/p_a$, so these equations cannot be inverted
to determine the two momenta separately.

The partially gauge-fixed theory nevertheless retains a residual gauge
redundancy, generated by the first-class constraint
$C_{\rm res}$. On the surface $\pi_N=0$, its generator
reduces to $N_{g.f.}\cH$. To exhibit the corresponding
transformation, let $\varepsilon_0(t)$ denote a parameter
for the reduced flow generated by $N_{g.f.}\cH$, and define
\begin{equation}
\delta_{\varepsilon_0}F
=
\varepsilon_0(t)\{F,N_{g.f.}\cH\}.
\end{equation}
Then
\begin{equation}
\delta_{\varepsilon_0}\mathcal L^{\rm red}
=
\dot\varepsilon_0\,N_{g.f.}\cH.
\end{equation}
This is a total derivative when $\varepsilon_0$ is constant,
as required for the residual symmetry. This constant
parameter should be distinguished from the original
time-dependent gauge parameter $\epsilon(t)$, whose
restriction to the gauge-fixing surface is $\epsilon(t)=\varepsilon_0N_{g.f.}(t)$.

We finally complete the physical reduction by fixing the residual freedom with $a=t$ and solving the Hamiltonian constraint. This second gauge
condition is transversal on either of the two branches
where $\dot a\neq0$. Solving $\cH=0$ for $p_a$ and
substituting into the reduced action gives
\begin{equation}
\mathcal L^{\rm red}\big|_{a=t,\,\cH=0}
=
p_\Phi\dot\Phi
\pm\sqrt{\frac{3}{4\pi G}}\frac{p_\Phi}{t}.
\label{eq:FLRW_canonical_rel_H}
\end{equation}
The sign $\pm$ distinguishes the two branches obtained
when solving $\cH=0$ for $p_a$.

Recall that from the covariantly dressed Lagrangian we obtained
\begin{equation}
H^\dr
=
\frac{N_\dr}{2t^3}(P_\Phi^\dr)^2
+\frac{3u}{8\pi GN_\dr}.
\end{equation}
Solving the residual dressed constraint
$-\cH_\dr=\cE^N_\dr\approx0$ gives
\begin{equation}
0\approx
\frac{1}{2u^3}(P_\Phi^\dr)^2
-\frac{3u}{8\pi G(N_\dr)^2},
\end{equation}
or
\begin{equation}
N_\dr
=
\pm\sqrt{\frac{3}{4\pi G}}\frac{u^2}{P_\Phi^\dr}.
\end{equation}
Substituting this into $H^\dr$ gives
\begin{equation}
H^\dr
=
\pm\sqrt{\frac{3}{4\pi G}}\frac{P^\dr_\Phi}{u}.
\end{equation}
This is the same branch-dependent function that appears in
the reduced action \eqref{eq:FLRW_canonical_rel_H}, up to
the identification of the dressed phase space with the gauge-fixed one and the identification of the gauge-fixed time $t$ with the relational time $u$.

The two procedures therefore agree at the classical level
on the appropriate constraint surface and branch, but they
do not produce the same intermediate variational
description. The covariantly dressed Lagrangian retains
the dressed lapse and its associated second-class
constraint pair. The canonical reduced description has
already fixed the lapse and eliminated the corresponding
constraints. In particular, performing a Legendre transform
of the canonically reduced Lagrangian does not reproduce
the covariantly dressed Lagrangian
\eqref{eq:FLRW_dressed_a_Lagr}.

The classical agreement established here does not by itself
prove quantum equivalence. In particular, one must still
determine whether the path integral for the covariantly
dressed theory, including its measure, contour, and treatment
of the dressed lapse, imposes the dressed constraint in the
same way as canonical reduction. Establishing this would
require an explicit path-integral quantization and a comparison
of the resulting physical inner product and transition
amplitudes with those of the reduced theory. We leave this
question open rather than assume that integrating over the
dressed lapse automatically reproduces the reduced quantum
theory.

\section{Conclusion}

We have investigated how covariant relational dressing, canonical constraint reduction, and relational Hamiltonian evolution are related in flat FLRW minisuperspace with a homogeneous massless scalar field. The model provides a setting in which these constructions can be compared explicitly, while retaining the essential features of a constrained gravitational system: time-reparametrization invariance, a Hamiltonian constraint, and dynamical variables that can serve as internal reference frames. Our analysis shows that the choice of relational frame affects not only how the fields are parametrized, but also the variational principle and the constraint structure of the resulting theory.

A lapse condition such as $N=1$, imposed at the level of the action, removes the independent lapse equation from the variational principle. The Noether identity \eqref{eq:FLRW_relationship} then implies only that the discarded constraint is constant along solutions, not that it vanishes. By contrast, suitable dynamical frames, such as $a$, $\Phi$ or even $\pi_a$, allow the remaining equations to recover the omitted frame equation through the off-shell Noether identity. The admissibility of a relational frame therefore requires more than the non-degeneracy of the frame map: its dressed variational principle must also retain enough independent equations to reproduce the original constrained solution space. This distinction explains why gauge fixing the lapse and constructing a relational dressing, despite their possible agreement on particular solutions, are not interchangeable operations.

The relational description also clarifies the status of its Hamiltonian. The dressed theory admits an integrable generator of evolution with respect to relational time, even when the corresponding transformations are not symmetries of the dressed action. In the $a$-frame, explicit dependence on relational time makes the Hamiltonian non-conserved, whereas in the $\Phi$-frame the absence of explicit time dependence yields a conserved generator associated with relational time translations. Thus, integrability of the generator does not require a corresponding Noether symmetry, and conservation follows only when the dressed theory possesses the additional symmetry. In both cases, the canonical analysis identifies the generator with the complete observable associated with the momentum conjugate to the chosen frame, on the corresponding solution branch. The comparison with the covariant construction nevertheless requires care: the canonical complete observable and the covariantly dressed momentum are distinct objects, related through the Legendre map and relational evaluation. Their identification with the relational Hamiltonian holds only after imposing the Hamiltonian constraint. The resulting generator is a bulk quantity arising from the relational Hamiltonian, rather than a charge associated with a spatial boundary or corner.

The distinction between gauge evolution and relational evolution is also reflected in the constraint structure. Time reparametrizations identify different parametrizations of the same history, whereas the relational Hamiltonian generates evolution relative to the chosen dynamical frame. Once the frame is fixed, there is no residual time-reparametrization gauge freedom in the relational description, although the dressed lapse continues to enforce a constraint. In the extended canonical formulation, the primary constraint associated with the lapse momentum and the secondary dressed Hamiltonian constraint form a second-class pair. They restrict the allowed relational phase space and determine the dressed lapse, rather than generating residual gauge transformations. The resulting relational dynamics is therefore governed by a non-vanishing Hamiltonian generator, which is not necessarily conserved in time.

Finally, covariant dressing and canonical reduction yield compatible classical descriptions on the appropriate constraint surface and solution branches, but they do not produce identical intermediate variational principles or phase spaces. The covariantly dressed theory retains the dressed lapse and its associated second-class constraints, whereas canonical reduction eliminates these structures through gauge-fixing and constraint reduction. In particular, the agreement of the relational generators on the constraint surface should not be interpreted as an off-shell equality. The correspondence is also branch-dependent and requires care wherever the rescaling of the constraint or the clock condition becomes singular. Whether the covariantly dressed path integral, with its measure, integration contour, and treatment of the dressed lapse specified, reproduces the physical inner product and transition amplitudes of reduced quantization remains an open question. The classical correspondence established here is therefore not, by itself, sufficient to establish quantum equivalence.

Several directions remain open. One concerns the relation between relational dressing and hidden or Schrödinger-type symmetries of cosmological models \cite{BenAchour:2019ywl,BenAchour:2022fif,Geiller:2022baq}. In particular, the conformal properties of the dressed action under rescalings of relational time may provide a starting point for identifying additional symmetries and their associated Hamiltonian generators. Another direction concerns the unitarity and equivalence of quantum theories constructed using different internal clocks, where the treatment of the lapse and of off-shell configurations may become essential. More broadly, extending the present analysis beyond minisuperspace to include inhomogeneous perturbations would allow us to investigate how the choice of relational frame affects the correlations and power spectra of cosmological perturbations. These questions require going beyond the homogeneous classical setting considered here.

\begin{acknowledgments}
The author would like to thank Sergio Aguilar-Gutierrez, Gonçalo Araújo-Regado, and Philipp Höhn for discussions on a related project \cite{JTproj} that provided inspiration for this work, as well as for their valuable insights and comments on the present manuscript. The author is also grateful to Ben Knepper and Luca Marchetti for discussions on path-integral applications of this work.
We acknowledge the forthcoming work \cite{AFHMtime}, which develops a generalisation of the framework presented in this paper.
The author is supported by the Okinawa Institute of Science and Technology Graduate University. This work was made possible through the support of the WOST, WithOut SpaceTime project (\hyperlink{https://withoutspacetime.org}{https://withoutspacetime.org}), supported by Grant ID\# 63683 from the John Templeton Foundation (JTF). The opinions expressed in this work are those of the author and do not necessarily reflect the views of the John Templeton Foundation.
\end{acknowledgments}

\appendix

\section{Covariant phase space and relational dressing}
\label{app:covariant phase space}

We briefly review the covariant phase-space construction underlying relational dressing, with particular emphasis on the variational principles relevant to the main text \cite{Carrozza:2022xut,Goeller:2022rsx,AFHMtime}. Consider a covariant theory with Lagrangian $L$ and independent fields $\psi^I$. Its variation takes the form
\begin{equation}
\delta L
=
E_I\delta\psi^I+\d\theta,
\end{equation}
where $E_I=0$ are the equations of motion and $\theta$
is the presymplectic potential.\footnote{We use $E_I$ for the Euler--Lagrange $d$-form and $\cE_I$ for its component density, such that $E_I = \cE_I\,\d^d x$ in a $d$-dimensional spacetime.}\\

\paragraph{Relational dressing and Noether identities.}

A reference frame $R$ defines a map from spacetime $\mathcal M$
to relational spacetime $\mathfrak m$. For geometric frames, the
dressing map can be chosen so that the dressed metric satisfies
a prescribed dressing condition. An example is the vanishing
of the component of the dressed connection along the direction
of a Wilson line, as in Wilson-line dressing in electromagnetism. The field-space Maurer--Cartan form associated with the frame is
\begin{equation}
\varpi:=\delta R^{-1}\circ R,
\qquad
\text{or, in coordinates,}\qquad
\varpi^\mu
=
-\frac{\partial x^\mu}{\partial R^A}\delta R^A.
\end{equation}
We introduce the horizontal variation
\begin{equation}
\delta_\varpi:=\delta+\pounds_\varpi,
\end{equation}
where the Lie derivative $\pounds_\varpi$, when acting on spacetime
forms, can be written as
$\pounds_\varpi=\d\iota_\varpi+\iota_\varpi\d$, for $\iota$ the spacetime contraction between vector and forms. We denote $\mathcal R$ the pull-back of forms from spacetime to relational spacetime.
The variation of a pulled-back form then satisfies
\begin{equation}
\delta\bigl(\mathcal R[\alpha]\bigr)
=
\mathcal R[\delta_\varpi\alpha].
\label{eq:fundamental_commutation}
\end{equation}

Let $\hat\xi$ denote the vector in field space associated with
a spacetime diffeomorphism generated by $\xi$. In particular, if the frame fields
$R^A$ transform as spacetime scalars,
\begin{equation}
\hat\xi\cdot\delta\psi^I=\pounds_\xi\psi^I,
\qquad
\hat\xi\cdot\varpi=-\xi.
\end{equation}

To characterise the covariance properties of
the presymplectic potential and the Lagrangian, let
$\mathrm L_{\hat\xi}$ denote the Lie derivative in field space
along $\hat\xi$. Following \cite{Carrozza:2022xut}, we define
the anomaly operator
\begin{equation}
\Delta_\xi
=
\mathrm L_{\hat\xi}
-\pounds_\xi
-\widehat{\delta\xi}\cdot,\label{eq:anomaly operator}
\end{equation}
which acts on differential forms as
\begin{equation}
\Delta_\xi
=
\delta\hat\xi\cdot
+\hat\xi\cdot\delta
-\d\iota_\xi
-\iota_\xi\d
-\widehat{\delta\xi}\cdot.
\end{equation}
We assume that
\begin{equation}
\Delta_\xi\theta=0,
\qquad
\Delta_\xi L=0,
\qquad
(\mathrm L_{\hat\xi}-\pounds_\xi)L=0.
\end{equation}
These conditions ensure an anomaly-free
presymplectic current and a covariant, anomaly-free Lagrangian, as is the case for the minisuperspace theory considered here.

The Noether current associated with the corresponding
diffeomorphism is
\begin{equation}
J_\xi
:=
\hat\xi\cdot\theta-\iota_\xi L
=
\mathcal C_\xi+\d q_\xi,
\qquad
\d J_\xi\approx0,
\label{eq:noether_current}
\end{equation}
where $\mathcal C_\xi\approx0$ is the constraint and $q_\xi$
is the Noether charge aspect. The corresponding off-shell
Noether identity is
\begin{equation}
\hat\xi\cdot\bigl(E_I\delta\psi^I\bigr)
+\d\mathcal C_\xi=0.
\label{eq:general_eom_relationship}
\end{equation}
For our model in this work, it reduces exactly to \eqref{eq:FLRW_relationship} with $C_\xi= -\xi \cE_N$. This identity is important when choosing dynamical fields as
relational coordinates. Since the frame fields become coordinates
on $\mathfrak m$, they are not varied in the standalone relational
variational principle. Their equations of motion must therefore
follow from the Noether identities if the dressed theory is to
reproduce the original dynamics \cite{Motohashi:2016prk}.
We now make this requirement more explicit,
first for general frame fields and then for the scalar frames
used in the main text.

Consider a $d$-dimensional covariant theory
written in terms of $M$ independent field components $\psi^I$.
For instance, if the theory contains a metric and $n$ independent
matter-field components, we have
$M=d(d+1)/2+n$ fields. The fields $\psi^I$ need not
be scalars: they can be independent components of tensorial or
spinorial matter fields, as well as metric components.
The diffeomorphism transformation of a general field component
can be written as
\begin{equation}
\hat\xi\cdot\delta\psi^I
=
\pounds_\xi\psi^I
=
\sum_{k=0}^{r_I-1}
\Delta^{I(k)}_\nu\partial_{(k)}\xi^\nu,
\end{equation}
where $r_I$ is the highest differential order
of the diffeomorphism parameter $\xi^\nu$ appearing in the
transformation of $\psi^I$, and $(k)$ denotes the multi-index
$(k)=\mu_1\cdots\mu_k$, with
$\partial_{(k)}=\partial_{\mu_1}\cdots\partial_{\mu_k}$.
The coefficients $\Delta^{I(k)}_\nu$ can depend on all the
components of the tensor to which $\psi^I$ belongs, and not
only on the component whose variation is being considered.
For scalar fields, by contrast, $r_I=1$ and
$\Delta^{I(0)}_\nu=\partial_\nu\psi^I$.

The variational principle gives
\begin{equation}
\delta L
=
\cE_I\delta\psi^I+\d\theta.
\end{equation}
In terms of the component densities $\cE_I$,
the Noether identity \eqref{eq:general_eom_relationship}
can then be rewritten as \cite{Motohashi:2016prk}
\begin{equation}
\xi^\nu
\sum_{I=1}^{M}\sum_{k=0}^{r_I-1}
(-1)^k\partial_{(k)}
\left(
\cE_I\Delta^{I(k)}_\nu
\right)
=0.
\end{equation}
Since $\xi^\nu$ is arbitrary, this implies
\begin{equation}
\sum_{I=1}^{M}\sum_{k=0}^{r_I-1}
(-1)^k\partial_{(k)}
\left(
\cE_I\Delta^{I(k)}_\nu
\right)
=0,
\qquad
\nu=1,\ldots,d.
\end{equation}

Suppose now that we identify $d$ independent
field components with the frame fields $R^A$, which become
coordinates on relational spacetime. Their equations of motion
$\cE_A=0$ do not follow directly from the variational principle
for $L^\dr$, since the frame coordinates are held fixed.
However, on shell of the equations of motion for the remaining
$M-d$ fields (equivalently, on shell of the corresponding
dressed equations of motion), the Noether identities reduce to
\begin{equation}
\sum_{A=1}^{d}\sum_{k=0}^{r_A-1}
(-1)^k\partial_{(k)}
\left(
\cE_A\widetilde\Delta^{A(k)}_\nu
\right)
\approx0,
\qquad
\nu=1,\ldots,d,
\label{eq:residual_eoms}
\end{equation}
where $\widetilde\Delta^{A(k)}_\nu$ denotes
the corresponding coefficients evaluated on shell of the
equations of motion for the other fields. These are, in general,
$d$ differential equations for the $d$ remaining equations
of motion $\cE_A$.

To ensure consistency with the original theory, for which
$\cE_A\approx0$ follows from the original variational principle,
we require that the only solution to
\eqref{eq:residual_eoms} be $\cE_A\approx0$.
Otherwise, the dressed theory may admit spurious solutions
that do not correspond to dressings of solutions of the
original theory. There are different ways to ensure this:
\begin{enumerate}
    \item If only the first order $k=0$ is non vanishing for each frame $R^A$ and the matrix $\widetilde\Delta^A_\nu$
    is non-degenerate, then \eqref{eq:residual_eoms} reduces to
    \begin{equation}
    \cE_A \widetilde\Delta^A_\nu \approx0,
    \end{equation}
    which implies $\cE_A\approx0$. By diagonalising $\Delta_\mu^A$ and opportunely rescaling the fields, this case is equivalent to having $d$ scalar frames.

    If at least one of the $R^A$ is
    not a scalar, the corresponding differential operator
    $\widetilde\Delta$ should be made invertible by
    imposing boundary conditions or restrictions on the initial
    data (if the differential equations involve derivatives
    along an on-shell time direction) such that the only
    solution is $\cE_A\approx0$.
\end{enumerate}

Conversely, if $\widetilde\Delta$ is not
invertible, spurious solutions can arise in the dressed theory.
Accordingly, the second possibility should be read as a sufficient condition once the relevant operator problem is well posed, not as a general theorem that every tensorial frame admits a locally or globally invertible dressing map. Anyway, dealing with non-scalar frames makes the covariant analysis considerably more involved, particularly when extracting information about the phase space and its symmetries. A canonical treatment in terms of gauge-fixing can then be more convenient \cite{JTproj}.\\

\paragraph{The two variational principles.}

There are two distinct ways to formulate the variational
principle in relational variables. The first is to pull back the original variational principle to relational spacetime, retaining variations of all fields,
including the frame coordinates. Using
\eqref{eq:fundamental_commutation}, we obtain
\begin{equation}
\delta\mathcal R[L]
=
\mathcal R[E_I\delta\psi^I]
+\d\mathcal R[\theta_\varpi],
\qquad
\theta_\varpi:=\theta+\iota_\varpi L.
\end{equation}
The corresponding presymplectic current is
\begin{equation}
\begin{aligned}
\omega_\varpi
&:=\delta_\varpi\theta_\varpi\\
&=
\omega-\iota_\varpi(E_I\delta\psi^I)
+\d\left(
\iota_\varpi\theta
+\frac12\iota_\varpi\iota_\varpi L
\right),
\end{aligned}
\label{eq:omega_chi}
\end{equation}
where $\omega:=\delta\theta$ is the original presymplectic
current. In particular, the pulled-back presymplectic current
differs from the original one by a term proportional to the
equations of motion and an exact form. The former vanishes
on shell, while the latter can contribute at corners if
spatial boundaries are present.

The second construction treats the pulled-back Lagrangian
\begin{equation}
L^\dr:=\mathcal R[L]
\end{equation}
as a standalone theory on fixed relational coordinates.
Writing the remaining dressed fields as
\begin{equation}
\Psi^i(R)
:=
\mathcal R[\psi^i],
\qquad i=1,\ldots,M-d,
\end{equation}
we vary only $\Psi^i$, keeping $R^A$ fixed. We denote this
variation by $\bar\delta$, so that
\begin{equation}
\bar\delta L^\dr
=
E_i^\dr\bar\delta\Psi^i
+\d\theta^\dr.
\label{eq:dressed_variation}
\end{equation}
Here $E_i^\dr$ and $\theta^\dr$ are respectively the
Euler--Lagrange forms and presymplectic potential of the
standalone relational theory.

The distinction between the two constructions is that the
pullback of the original variational principle also accounts
for variations of the frame coordinates, whereas the standalone
relational variational principle does not. This difference can
be expressed in terms of the constraint. With the standard
Noether-charge improvement of the presymplectic potential,
define
\begin{equation}
\Theta
:=
\mathcal R[\theta_\varpi+\d q_\varpi].
\end{equation}
Using \eqref{eq:noether_current}, one finds
\begin{equation}
\Theta
=
\theta^\dr-\mathcal R[\mathcal C_\varpi].
\label{eq:theta_difference}
\end{equation}
The second term is the dressed constraint contracted with
the field-space variation of the frame.

Equation \eqref{eq:theta_difference} makes explicit that the
two presymplectic potentials need not agree off shell. Their
difference vanishes on the physical constraint surface,
up to the boundary contributions associated with the
Noether-charge improvement. Thus, agreement of the physical
presymplectic structures does not imply equality of the
off-shell variational principles.

This distinction is relevant when comparing covariant dressing
with canonical reduction. The covariant construction retains
the dressed constraint and the frame variations before
restricting to the physical phase space, whereas canonical
reduction eliminates the frame variables by imposing the
corresponding constraints and gauge conditions. The two
descriptions can therefore yield the same relational dynamics
while having different intermediate variational and
presymplectic structures.

\section{The scalar-field frame.}
\label{app:scalar}

The discussion in the main text applies equally well when $R=\Phi$. In this case, the
dressed Lagrangian is independent of relational time. The
covariant variation under a relational transformation is
\begin{equation}
\hat\rho\cdot\bar\delta L_\dr
=
-\rho\,\partial_t(N_\dr\cE^N_\dr)
-\dot\rho
\underbrace{\left(
\frac{a_\dr^3}{2N_\dr}
+\frac{3}{8\pi G}
\frac{a_\dr\dot a_\dr^2}{N_\dr}
\right)}_{\not\approx\,0}.
\end{equation}
The term proportional to $\dot\rho$ is not zero off shell.
Nevertheless, for constant $\rho$, the dressed action is
invariant under relational time translations, and the
corresponding evolution generator is conserved on shell.

The dressed presymplectic structure is
\begin{equation}
\mathcal R_\Phi[\omega_\varpi]
=
\delta\,\pi_a^\dr\curlywedge\delta a_\dr
+
\mathcal R_\Phi\left[
\frac{\delta\Phi}{\dot\Phi}
\right]
\curlywedge\delta(N_\dr\cE^N_\dr).
\end{equation}
On the physical constraint surface, the second term drops
out. The relational Hamiltonian is
\begin{equation}
H_\Phi
=
\frac{a_\dr^3}{2N_\dr}
+\frac{3}{8\pi G}
\frac{a_\dr\dot a_\dr^2}{N_\dr}.
\end{equation}
Using the constraint, this becomes
\begin{equation}
H_\Phi
\approx
\frac{a_\dr^3}{N_\dr}
=
\pi_\Phi^\dr,
\end{equation}
and is conserved on shell.

In the canonical description, the same result follows
immediately:
\begin{equation}
\cO_{p_\Phi}^{\Phi_0}
=
\int\d t\,
\delta(\Phi_0-\alpha_t(\Phi))\,
\alpha_t(p_\Phi)\,
\alpha_t(\{\Phi,X\})
=
p_\Phi.
\end{equation}
Since $\pi_\Phi=a^3\dot\Phi/N$ and $\dot\Phi=1$ in the
relational description, we obtain
\begin{equation}
\cO_{p_\Phi}^{\Phi_0}
\longleftrightarrow H_\Phi.
\end{equation}

The two frames therefore exhibit the same canonical
correspondence between the momentum conjugate to the frame
and the generator of relational evolution. Their difference
lies instead in the symmetry properties of the dressed
action: the generator in the $a$-frame is not conserved,
whereas in the $\Phi$-frame it is conserved on shell.

\section{Equations of motion for the momentum frame}
\label{app:no_sol}

We show that the two equations \eqref{eq:pi_phi_eoms} do not admit simultaneous solutions. Assuming $a\neq0$,
$\dot\Phi\neq0$ and $G\neq0$, they can be rewritten as
\begin{align}
\f{4\pi G a \dot \Phi^2}{3}\cE_a^\dr = &u\ddot a\dot\Phi
+\dot a(\dot\Phi-u\ddot\Phi)
-\frac{u\dot a^2}{a}\dot\Phi \approx 0\,,
\label{eq:app_first}\\
8\pi G a^3 \dot \Phi^3\cE_\Phi^\dr =&6u\dot a^3\dot\Phi
-3a\dot a\left(
2u\ddot a\dot\Phi
+\dot a(\dot\Phi-2u\ddot\Phi)
\right)
-4\pi G a^3\dot\Phi^3\approx 0
\label{eq:app_second}
\end{align}

Introducing $\zeta=\dot a/a$, the first equation becomes
\begin{equation}
u\dot\Phi\dot \zeta+\zeta(\dot\Phi-u\ddot\Phi)\approx 0,
\end{equation}
which can be integrated to give
\begin{equation}
 \zeta \approx C\frac{\dot\Phi}{u},
\label{eq:app_H_solution}
\end{equation}
where $C$ is an integration constant.

On the other hand, rewriting the second equation in terms of
$\zeta$ and using the first equation, we obtain
\begin{equation}
3\zeta^2\dot\Phi-4\pi G\dot\Phi^3\approx 0.
\end{equation}
Since $\dot\Phi\neq0$, this implies
\begin{equation}
\zeta ^2=\frac{4\pi G}{3}\dot\Phi^2.
\end{equation}
Substituting \eqref{eq:app_H_solution}, we are left with
\begin{equation}
\frac{3C^2}{4\pi G u^2}=1.
\end{equation}
This condition cannot hold identically on any open interval
of the independent variable $u$, since $C$ and $G$ are
constants. Hence, the two equations are incompatible and
the system admits no nondegenerate solutions.

\renewcommand{\refname}{References} 
\phantomsection
\bibliographystyle{biblio}
\bibliography{references}
\end{document}

%% file: fieldspace_original.tex

\begin{tikzpicture}[x=0.025cm, y=-0.025cm, >=Stealth,every node/.style={font=\sffamily}]
  \tikzset{bigarrow/.style={-{Triangle[length=6.4mm,width=6.4mm]}, line width=3.4pt, ink},
           curve/.style={arr, line width=2.7pt, -{Stealth[length=5.1mm,width=4.8mm]}}}

  \coordinate (FL) at (110,600);
\coordinate (FR) at (612,590);
\coordinate (BL) at (336,455);
\coordinate (BR) at (828,455);
\coordinate (TL) at (110,348);
\coordinate (TM) at (336,203);
\coordinate (TR) at (828,203);
\foreach \i in {0,0.1,...,1.001}
    \draw[grid, line width=0.8pt] ($(FL)!\i!(FR)$) -- ($(BL)!\i!(BR)$);
\foreach \j in {0,0.167,...,1.001}
    \draw[grid, line width=0.8pt] ($(FL)!\j!(BL)$) -- ($(FR)!\j!(BR)$);
\foreach \j in {0,0.1,...,1.001}
    \draw[grid, line width=0.8pt] ($(BL)!\j!(BR)$) -- ($(TM)!\j!(TR)$);
\foreach \j in {0,0.167,...,1.001}
    \draw[grid, line width=0.8pt] ($(FL)!\j!(BL)$) -- ($(TL)!\j!(TM)$);
\foreach \j in {0,0.15,...,1.001}
    \draw[grid, line width=0.8pt] ($(FL)!\j!(TL)$) -- ($(BL)!\j!(TM)$);
\foreach \j in {0,0.15,...,1.001}
    \draw[grid, line width=0.8pt] ($(BL)!\j!(TM)$) -- ($(BR)!\j!(TR)$);

  \draw[bigarrow] (115,600) -- (115,348);
\node[ink, font=\LARGE] at (86,358) {$a$};
\draw[bigarrow] (115,600) -- (652,590);
\node[ink, font=\LARGE] at (675,588) {$N$};
\draw[bigarrow] (115,600) -- (336,455);
\node[ink, font=\LARGE] at (268,448) {$\Phi$};

  \fill[band, opacity=0.8] (420,301.5) -- (620,286) -- (527.1,535.5) -- (327.1,540) -- cycle;
\fill[disc] (427.1,540) ellipse [x radius=100, y radius=25];
\draw[fib!60!ink, line width=2.4pt, opacity=0.7] (427.1,540) ellipse [x radius=100, y radius=25];

  \foreach \bx/\tx/\ty in {260/370/282, 312/440/285, 465/600/285, 540/670/285}
    \draw[fib, line width=8.5pt, line cap=round, opacity=0.75]
      (\bx,548) .. controls ({\bx-22},500) and ({\tx-45},{\ty+120}) .. (\tx,\ty);
\draw[fibd, line width=8.5pt, line cap=round]
      plot[smooth, tension=0.7] coordinates {(390,548) (402,505) (443,420) (463,353) (482,312) (510,286)};

  \fill[plane, opacity=0.6] (110,495) -- (611.75,495) -- (824.24,345) -- (310,355) -- cycle;
\draw[planeb, line width=2.7pt] (110,495) -- (611.75,495) -- (824.24,345) -- (310,355) -- cycle;

  \fill[disc] (520,286) ellipse [x radius=100, y radius=26];


 \draw[ink, line width=3.4pt, fill=white] (418,469.84) circle (7pt);
\draw[ink, line width=3.4pt, fill=cy]    (450,403) circle (7pt);
\draw[cy!70!tl, line width=2pt, -{To[length=2.5mm]}] (424,456) -- (443.28,416);

  \node[ink!85!black, anchor=west, font=\bfseries\LARGE] at (838,222) {Gauge fiber};
\node[head, anchor=north west, font=\large, align=left, text width=14cm] at (838,240)
    {};
\draw[curve] (480,340) .. controls (540,350) and (650,225) .. (815,220);
\node[ink, anchor=west, font=\Large] at (917,322) {Gauge fixing  $a=a_1(t)$};
\draw[curve] (745,418) .. controls (840,425) and (820,335) .. (908,335);
\node[ink, anchor=east, font=\large] at (733,660) {solution space (stationary subspace)};
\draw[ink, line width=1.2pt, -{Stealth[length=3.2mm]}] (480,640) -- (420,560);

  \draw[ink, line width=1.4pt, -{Stealth[length=3.2mm]}] (925.88,560.42) -- (1200.88,560.42);
\draw[ink, line width=1.4pt, -{Stealth[length=3.2mm]}] (925.88,559.5) -- (925.88,390.5);
\node[ink, font=\normalsize] at (1215,558) {t};
\node[ink, font=\normalsize, rotate=90] at (914.99,471.02) {fields};

\draw[pk, line width=3.7pt, domain=927:1200, samples=20, smooth, variable=\x]
  plot ({\x},{549 - (1/0.0321)*(((\x-927)*0.021-3)/2+2)});
\draw[fib, line width=3.7pt, domain=927:1200, samples=20, smooth, variable=\x]
  plot ({\x},{549 - (1/0.0321)*exp(((\x-927)*0.021-3)/3)});
\draw[nn, line width=3.7pt, domain=927:1150, samples=20, smooth, variable=\x]
  plot ({\x},{549 - (1/0.0321)*exp(((\x-927)*0.021-3))});

\node[pk, font=\Large] at (1208.8,440) {$\Phi$};
\node[nn, font=\Large] at (1168.96,380) {$N$};
\node[fib, font=\Large] at (1210.75,471.02) {$a$};
  \foreach \y in {540.501,528.796,506.928} \draw[ink, line width=2.4pt, fill=white] (1008,\y) circle (5pt);
  \foreach \y in {507.946,514.845,482.396} \draw[ink, line width=2.4pt, fill=cy]    (1083,\y) circle (5pt);
\draw[cy!40!tl, line width=1.5pt, -{To[length=2.5mm]}] (1008,492.66) -- (1080,467.6);
\node[ink, font=\normalsize] at (1058,572) {};

  \draw[curve] (1083,456) .. controls (951.5,388.14) and (701.5,383.14) .. (485.5,407.14);
\draw[curve] (462,471) .. controls (650,545) and (780,560) .. (906,541);

  \node[tl, anchor=west, font=\large] at (894.12,592.34) {Movement along fibers $\approx$ Movement along histories};
\node[tl, anchor=west, font=\large] at (918.12,628.4) {$\psi(t)\to \psi(t+\xi)$};
\end{tikzpicture}

%% file: fieldspace_dressed.tex


\begin{tikzpicture}[x=0.025cm, y=-0.025cm, >=Stealth]
  \tikzset{bigarrow/.style={-{Triangle[length=6.4mm,width=6.4mm]}, line width=3.4pt, ink},
           curve/.style={arr, line width=2.7pt, -{Stealth[length=5.1mm,width=4.8mm]}}}

  \coordinate (FL) at (110,450);
\coordinate (FR) at (612,450);
\coordinate (BL) at (336,305);
\coordinate (BR) at (828,305);
\coordinate (TL) at (110,198);
\coordinate (TM) at (336,53);
\coordinate (TR) at (828,53);
\foreach \i in {0,0.1,...,1.001}
    \draw[grid, line width=0.8pt] ($(FL)!\i!(FR)$) -- ($(BL)!\i!(BR)$);
\foreach \j in {0,0.167,...,1.001}
    \draw[grid, line width=0.8pt] ($(FL)!\j!(BL)$) -- ($(FR)!\j!(BR)$);
\foreach \j in {0,0.1,...,1.001}
    \draw[grid, line width=0.8pt] ($(BL)!\j!(BR)$) -- ($(TM)!\j!(TR)$);
\foreach \j in {0,0.167,...,1.001}
    \draw[grid, line width=0.8pt] ($(FL)!\j!(BL)$) -- ($(TL)!\j!(TM)$);
\foreach \j in {0,0.15,...,1.001}
    \draw[grid, line width=0.8pt] ($(FL)!\j!(TL)$) -- ($(BL)!\j!(TM)$);
\foreach \j in {0,0.15,...,1.001}
    \draw[grid, line width=0.8pt] ($(BL)!\j!(TM)$) -- ($(BR)!\j!(TR)$);

  \fill[disc] (444.84,396.92) ellipse [x radius=125, y radius=25];
\draw[fib!60!ink, line width=2.7pt, opacity=0.7] (444.84,400) ellipse [x radius=125, y radius=25];
\fill[band, opacity=0.8] (322,125) -- (572,131.21) -- (569.84,400) -- (322,400) -- cycle;
\foreach \x/\ty/\by/\dx in {352/115/390/-6, 252/125/410/-6, 420/115/400/-4, 520/125/392/-2,652/115/380/-3}
    \draw[fib, line width=8.5pt, line cap=round]
      (\x,\by) .. controls ({\x+\dx},{\by-100}) and ({\x-\dx},{\ty+90}) .. ({\x+\dx*0.8},\ty);
\draw[og, line width=2.0pt, -{To[length=2.5mm]}] (410.16,268.92) -- (369.47,239.68);
\draw[ink, line width=3.4pt, fill=white] (420,275) circle (7pt);

  \draw[ink, line width=3.4pt, fill=pe]    (355,229.84) circle (7pt);

  \draw[bigarrow] (110,450) -- (110,185);
\node[ink, font=\LARGE] at (73.78,214.79) {$a$};
  \draw[bigarrow] (110,450) -- (680,450);
\node[ink, font=\LARGE] at (727.52,450.08) {$N_{\rm dr}$};
\draw[bigarrow] (110,450) -- (336,305);
\node[ink, font=\LARGE] at  (304.44,294.92) {$\Phi_{\rm dr}$};
\fill[disc] (451.16,523) ellipse [x radius=128, y radius=25];
\draw[fib!60!ink, line width=2.7pt] (451.16,523) ellipse [x radius=128, y radius=25];

  \draw[og, line width=2.0pt, -{To[length=2.5mm]}] (419,521.5) -- (380,518.5);
\draw[ink, line width=3.4pt, fill=pe]    (364.71,520) circle (7pt);
\draw[ink, line width=3.4pt, fill=white] (430.16,523) circle (7pt);
\node[head, anchor=east, font=\Large] at (840.35,538.91) {relational fields: $\Phi_{\rm dr},\;N_{\rm dr}$};
\node[head, anchor=east, font=\Large] at (753.49,60.21) {only $a$ varies along the fibers};
\draw[curve] (430.16,305) .. controls (619.27,353.62) and (720,375) .. (909.98,362.38);

\draw[og, line width=1.5pt, -{To[length=2.5mm]}] (1000.97,277) -- (1075.97,225.64);
\fill[disc] (444.84,125) ellipse [x radius=125, y radius=25];
\draw[fib!60!ink, line width=2.7pt, opacity=0.7] (447,125) ellipse [x radius=125, y radius=25];
\draw[pe, line width=3pt, -{Stealth[length=5.0mm]}, xscale=-1] (-1014.99,240) .. controls (-759.96,119.68) and (-599.96,119.68) .. (-372.47,220);
\draw[ink, line width=1.4pt, -{Stealth[length=5.0mm]}] (347.19,377) .. controls (297.19,435) and (307.19,485) .. (344.19,520);

\draw[fib, line width=3.7pt, domain=927:1200, samples=10, smooth, variable=\x]
  plot ({\x},{350 - (1/0.0321)*((\x-927)*0.021)});
\draw[pk, line width=3.7pt, domain=940:1200, samples=30, smooth, variable=\x]
  plot ({\x},{350 - (1/0.0321)*2*ln((\x-927)*0.021)});
\draw[nn, line width=3.7pt, domain=927:1200, samples=30, smooth, variable=\x]
  plot ({\x},{350 - (1/0.0321)*0.15*(((\x-927)*0.021))^2});

\node[pk, font=\Large] at (1219.74,251.52) {$\Phi_\text{dr}$};
\node[nn, font=\Large] at (1227.23,200) {$N_\text{dr}$};
\node[fib, font=\Large] at (1210.75,160) {$a$};
  \foreach \y in {297.009,336.479,316.902} \draw[ink, line width=2.4pt, fill=white] (1008,\y) circle (5pt);
  \foreach \y in {247.944,299.85,276.067} \draw[ink, line width=2.4pt, fill=pe]    (1083,\y) circle (5pt);
\node[ink, font=\normalsize] at (1058,572) {};

  \draw[ink, line width=1.4pt, -{Stealth[length=3.2mm]}] (925.88,350) -- (1200.88,350);
\draw[ink, line width=1.4pt, -{Stealth[length=3.2mm]}] (925.88,450) -- (925.88,190.5);
\node[ink, font=\normalsize] at (1215,358) {t};
\node[ink, font=\normalsize, rotate=90] at (914.99,271.02) {fields};

\end{tikzpicture}
